\documentclass[aps,prb,reprint,superscriptaddress,nofootinbib,longbibliography,floatfix]{revtex4-2}

\usepackage{amsmath}
\usepackage{amssymb}
\usepackage{bm}
\usepackage{physics}
\usepackage{graphicx}
\usepackage[utf8]{inputenc}
\usepackage{xcolor}
\usepackage{array}[=2016-10-06]
\usepackage{tabularx}
\usepackage{multirow}
\usepackage[colorlinks=true,pdfstartview=FitV,linkcolor=blue,citecolor=blue,urlcolor=blue]{hyperref}

\newcommand{\Eq}[1]{Eq.~\eqref{#1}}
\newcommand{\Fig}[2][]{\hyperref[#2]{Fig.~\ref*{#2}#1}}
\newcommand{\Sec}[1]{Sec.~\ref{#1}}

\newcommand{\intk}{\int\!\frac{d^d k}{(2\pi)^d}}

\begin{document}

\title{Interaction-Induced Bloch Geometry and Transverse Thermoelectric Transport}

\author{Heng-Yu Di}
\affiliation{School of Physical Sciences, University of Chinese Academy of Sciences, Beijing 100049, China}

\author{YuanDong Wang}
\affiliation{Department of Applied Physics, College of Science, China Agricultural University, Beijing 100083, China}

\author{Zhen-Gang Zhu}
\email{zgzhu@ucas.ac.cn}
\affiliation{School of Physical Sciences, University of Chinese Academy of Sciences, Beijing 100049, China}
\affiliation{School of Electronic, Electrical and Communication Engineering, University of Chinese Academy of Sciences, Beijing 100049, China}

\author{Gang Su}
\email{gsu@ucas.ac.cn}
\affiliation{Institute of Theoretical Physics, Chinese Academy of Sciences, Beijing 100190, China}
\affiliation{School of Physical Sciences, University of Chinese Academy of Sciences, Beijing 100049, China}
\affiliation{Kavli Institute for Theoretical Sciences, University of Chinese Academy of Sciences, Beijing 100190, China}

\date{October 21, 2026}

\begin{abstract}
Electron--electron interactions can reconstruct Bloch wave functions, changing their geometry and the associated thermoelectric response.
We derive the linear thermoelectric response within the Hartree--Fock (HF) approximation using the thermal length-gauge density-matrix approach, resolving Drude, symmetric quantum-metric, and antisymmetric Berry-curvature contributions.
In a noninteracting Bloch eigenbasis, we separate the reconstructed Berry connection into the noninteracting connection and an interaction connection. This gives bare, coherence, and interaction-connection contributions to the anomalous thermoelectric response.
We determine the constraints imposed by spatial inversion, rotational, time-reversal, and combined symmetries on the three transport channels. In the relaxation-time approximation, the leading weak-broadening behavior is $\Gamma^{-1}$ for the Drude contribution, $\Gamma$ for the quantum-metric contribution, and $\Gamma^0$ for the Berry-curvature contribution. The interband scalings apply away from band degeneracies.
Self-consistent calculations for the triangular-lattice extended Hubbard model illustrate how interactions generate Berry curvature and a quantum metric, change the symmetry of the system, and thereby contribute to transverse thermoelectric transport.
\end{abstract}

\maketitle

\raggedbottom
\clubpenalty=10000
\widowpenalty=10000

\section{Introduction}
\label{sec:introduction}

The geometry of Bloch wave functions can influence thermoelectric transport in crystals through interband response.
Berry curvature underlies intrinsic anomalous Hall transport \cite{Nagaosa2010}, while the corresponding anomalous thermoelectric response is sensitive to the Berry-curvature distribution near the Fermi level \cite{Xiao2006,Xiao2010,Noky2018ANE,Minami2020NodalLines}.
The quantum metric can also contribute to linear thermoelectric response through interband coherence \cite{WenXieAuerbachUchoa2025,WenErratum2026,LhachemiCano2026}.
Electron--electron interactions can change Bloch eigenstates and thereby reconstruct their Berry curvature and quantum metric.

Studies of $\mathrm{Co}_2\mathrm{MnGa}$ \cite{Sakai2018Co2MnGa} and $\mathrm{Mn}_3\mathrm{Sn}$ \cite{Ikhlas2017Mn3Sn,Li2017Mn3Sn} show that the anomalous Nernst response is sensitive to the Berry-curvature distribution near the Fermi level.
Chemical-potential tuning in $\mathrm{Fe}_3\mathrm{Sn}_2$ \cite{Li2023Fe3Sn2} further illustrates this sensitivity.
In $\mathrm{Co}_3\mathrm{Sn}_2\mathrm{S}_2$, a large anomalous Nernst response was linked to Berry curvature near the Fermi level \cite{Guin2019Co3Sn2S2,Yang2020}, while subsequent optical measurements and many-body calculations revealed correlation-induced reconstruction of the electronic bands \cite{Xu2020CorrelatedCo3Sn2S2}.
For $\mathrm{UCo}_{0.8}\mathrm{Ru}_{0.2}\mathrm{Al}$, electronic correlations were further proposed to shift band crossings toward the Fermi level, bringing states with strong Berry curvature into the energy window relevant to thermoelectric transport \cite{Asaba2021}.
Hybridization-induced reconstruction of electronic states has been linked to enhanced anomalous Hall transport in $\mathrm{La}_{1-x}\mathrm{Ce}_x\mathrm{Co}_2\mathrm{As}_2$ \cite{Cheng2023KondoHall}.
The ferromagnetic Kondo systems $\mathrm{CeCo}_2\mathrm{As}_2$ and $\mathrm{CeCrGe}_3$ illustrate the role of hybridization: $f$--$d$ hybridization forms narrow bands with pronounced Berry curvature near the Fermi level while changing the orbital character and energy distribution of the electronic states \cite{Guan2026,Li2026}.

Thermal response can be formulated in terms of equilibrium correlation functions \cite{KuboYokotaNakajima1957}. Luttinger's thermal field provides a way to introduce thermal driving into this description \cite{Luttinger1964}. These methods have been used to study anomalous electric, thermoelectric, and thermal Hall transport \cite{Onoda2008}. Measurable transport coefficients require subtracting local magnetization currents from the current response \cite{Cooper1997,QinNiuShi2011,KapustinSpodyneiko2021}.

For interacting systems, transport has been studied through perturbative corrections and through the states generated by interactions.
Using diagrammatic perturbation theory, previous studies have evaluated first-order electron--electron interaction corrections to the anomalous Hall \cite{DumitriuIovanescu2024} and thermoelectric Hall \cite{DumitriuIovanescu2026} responses of two-dimensional massive Dirac fermions.
Studies of Hubbard-type models have investigated interaction-driven topological states arising from chiral magnetic order \cite{JiangZhangZhouWang2015}, interaction-generated complex hoppings \cite{PanWuDasSarma2020}, and inversion of a correlated charge-transfer gap accompanied by magnetic reconstruction \cite{DevakulFu2022}.
Mean-field studies have calculated Berry curvature from Bloch wave functions in chiral $d$-density-wave states and related it to the anomalous Nernst response \cite{ZhangTewariYakovenkoDasSarma2008}. Self-consistent HF studies have also examined how interaction-induced, temperature-dependent band renormalization affects charge and heat transport \cite{WangZhu2026}. These studies provide a basis for analyzing transport in correlated systems through mean-field eigenstates and their geometry.

We use the thermal length-gauge density-matrix approach within the self-consistent Hartree--Fock (HF) approximation to study linear thermoelectric response.
The diagonal density-matrix response gives the Drude contribution, while the off-diagonal response gives the quantum-metric and Berry-curvature contributions \cite{Tatara2015,Wang2022b}. We include orbital magnetization in the transport coefficients.
In the noninteracting Bloch eigenbasis, we separate the reconstructed Berry connection into the noninteracting connection and an interaction connection, thereby resolving the anomalous thermoelectric response into a bare geometric term, a coherence term coupling the two connections, and a term built solely from the interaction connection.
We further analyze the constraints imposed by spatial, time-reversal, and combined symmetries on the three transport channels and examine their broadening dependence in the relaxation-time approximation.
As a concrete example, we study the triangular-lattice extended Hubbard model, in which self-consistent onsite and bond mean fields reconstruct the Berry curvature and quantum metric of Bloch states and contribute to transverse thermoelectric transport through changes in geometry and symmetry.

The remainder of this paper is organized as follows.
Section~\ref{sec:mean-field-thermoelectric} derives the linear thermoelectric response of mean-field systems, and Section~\ref{sec:interaction-geometry} develops its geometric representation in the noninteracting basis.
Section~\ref{sec:scattering-symmetry} analyzes the symmetry and broadening characteristics of the transport channels, and Section~\ref{sec:triangular-results} illustrates these mechanisms in the triangular-lattice model.

\section{Thermoelectric response of interacting mean-field systems}
\label{sec:mean-field-thermoelectric}

\subsection{HF approximation}

We consider a translationally invariant tight-binding model formulated in a localized-orbital basis.
The operator $c_{\bm R\lambda}$ annihilates an electron in orbital $\lambda$ of unit cell $\bm R$, whose physical position is $\bm r_{\bm R\lambda}=\bm R+\bm\tau_\lambda$.
The noninteracting tight-binding Hamiltonian is given by
\begin{equation}
H_0=\sum_{\bm R,\bm R'}\sum_{\lambda,\lambda'}
t_{\lambda\lambda'}(\bm R-\bm R')
c_{\bm R\lambda}^{\dagger}c_{\bm R'\lambda'},
\label{eq:real-space-H0}
\end{equation}
where $t_{\lambda\lambda'}$ denotes the hopping amplitude from orbital $\lambda'$ in cell $\bm R'$ to orbital $\lambda$ in cell $\bm R$.
The general two-particle interaction takes the form
\begin{equation}
\begin{aligned}
H_{\mathrm{int}}&=\frac{1}{2}
\sum_{\{\bm R_i,\lambda_i\}_{i=1}^{4}}
V_{1234}(\{\bm R_i,\lambda_i\})\\
&\quad\times
c_{\bm R_1\lambda_1}^{\dagger}c_{\bm R_2\lambda_2}^{\dagger}
c_{\bm R_3\lambda_3}c_{\bm R_4\lambda_4},
\end{aligned}
\label{eq:real-space-Hint}
\end{equation}
with $V_{1234}$ representing the interaction matrix element among four localized orbitals.
Performing a Fourier transformation to momentum space, $c_{\bm k\lambda}=N_{\mathrm c}^{-1/2}\sum_{\bm R}e^{-i\bm k\cdot\bm R}c_{\bm R\lambda}$, where $N_{\mathrm c}$ is the number of unit cells, we obtain the noninteracting Bloch Hamiltonian and its eigenstates $\ket{a(\bm k)}$.
The unitary eigenvector matrix $u_{a\lambda}(\bm k)=\langle a(\bm k)|\lambda(\bm k)\rangle$ transforms the orbital operators to the noninteracting band basis through $c_{\bm k a}=\sum_\lambda u_{a\lambda}(\bm k)c_{\bm k\lambda}$.
In this basis, the noninteracting Hamiltonian is diagonal, $H_0=\sum_{\bm k,a}\epsilon_{\bm k a} c_{\bm k a}^{\dagger}c_{\bm k a}$.
Because the bare Hamiltonian is diagonal, all off-diagonal matrix elements of the mean-field Hamiltonian in this basis originate from the HF self-energy.

We decouple the interaction in the HF approximation (Appendix~\ref{app:hartree-fock-details}).
The translationally invariant mean-field one-particle density matrix $\rho_{ab}(\bm k)=\langle c_{\bm k b}^{\dagger}c_{\bm k a}\rangle$ contains both band occupations and interband coherences.
The HF decoupling yields the effective Hamiltonian
\begin{equation}
H_{\mathrm{MF}}=E_{\mathrm{const}}[\rho]
+\sum_{\bm k,a,b}[h_{\mathrm{MF}}(\bm k)]_{ab}
c_{\bm k a}^{\dagger}c_{\bm k b},
\label{eq:general-HMF}
\end{equation}
where $E_{\mathrm{const}}[\rho]$ is the $c$-number energy offset generated by the decomposition.
The renormalized single-particle matrix is
\begin{equation}
[h_{\mathrm{MF}}(\bm k)]_{ab}
=\epsilon_{\bm k a}\delta_{ab}
+\Sigma^{\mathrm H}_{ab}(\bm k)+\Sigma^{\mathrm F}_{ab}(\bm k),
\label{eq:mean-field-single-particle-matrix}
\end{equation}
Here, $\Sigma^{\mathrm H}$ denotes the Hartree self-energy and $\Sigma^{\mathrm F}$ the Fock self-energy.
The unitary transformation $c_{\bm k a}=\sum_n W_{an}(\bm k)d_{\bm k n}$ diagonalizes the single-particle matrix in the interacting quasiparticle basis, giving
\begin{equation}
H_{\mathrm{MF}}=E_{\mathrm{const}}
+\sum_{\bm k,n}\mathcal E_{\bm k n}
d_{\bm k n}^{\dagger}d_{\bm k n}.
\label{eq:diagonal-mean-field-Hamiltonian}
\end{equation}
The interacting eigenstates are denoted by $\ket{n(\bm k)}$, with $W_{an}(\bm k)=\langle a(\bm k)|n(\bm k)\rangle$. The equilibrium occupations are given by the Fermi--Dirac distribution, $f_{\bm k n}=f(\mathcal E_{\bm k n}-\mu)$, where $f(x)=(e^{\beta x}+1)^{-1}$, $\beta=1/(k_{\mathrm B}T)$, and $\mu$ is the chemical potential. In the discussion of band spectra, the Fermi level denotes the equilibrium chemical potential at the specified temperature and filling.
The self-consistency condition in the noninteracting basis is $\rho_{ab}(\bm k)=\sum_n W_{an}(\bm k)W_{bn}^{*}(\bm k)f_{\bm k n}$.

\subsection{Thermal driving and density-matrix response}

We calculate the linear response of the equilibrium HF Hamiltonian at the chosen temperature, keeping the mean fields fixed.

We define $\widetilde h_{\mathrm{MF}}=h_{\mathrm{MF}}-\mu I$ as the single-particle mean-field Hamiltonian measured relative to the chemical potential. Luttinger's thermal field $\bm E_T=-\bm\nabla T/T$ couples through the Hermitian length-gauge perturbation $h_T(t)=-\frac{1}{2}\{\widetilde h_{\mathrm{MF}},\bm r\cdot\bm E_T(t)\}$.
In the momentum representation of the interacting basis, the position operator has the covariant-derivative form $[r_{\bm k}^{\mu}]_{mn}=i\delta_{mn}\partial_{k_\mu}+\mathcal A_{mn}^{\mu}$, where $\mathcal A_{mn}^{\mu}=i\langle m(\bm k)|\partial_{k_\mu}|n(\bm k)\rangle$ is the interacting Berry connection.
For brevity in the perturbation term, we define the anticommutator $q_{\bm k}^{\mu}=\{\widetilde h_{\mathrm{MF}},r_{\bm k}^{\mu}\}$.

Let $\varrho(t)$ denote the many-body density operator in the interacting mean-field Fock space.
The corresponding one-particle reduced density matrix is defined as $\rho_{mn}(\bm k,t)=\operatorname{Tr}[d_{\bm k n}^{\dagger}d_{\bm k m}\varrho(t)]$.
Under the total Hamiltonian $H(t)=H_{\mathrm{MF}}+H_T(t)$, the reduced density matrix obeys the Liouville--von Neumann equation, $i\hbar\partial_t\rho_{mn}(\bm k,t)=\langle[d_{\bm k n}^{\dagger}d_{\bm k m},H(t)]\rangle$.
We expand the density matrix in powers of the thermal field, $\rho=\sum_{p\geq0}\rho^{(p)}$, and Fourier transform to the frequency domain. The resulting recursion relation is (Appendix~\ref{app:density-matrix-details})
\begin{equation}
    \begin{aligned}
    \rho_{mn}^{(p+1)}(\bm k,\omega)
    =-&\frac{1}{2(\hbar\omega-\mathcal E_{\bm k,mn}+i\eta)}
    \int\frac{d\omega_1}{2\pi}E_T^\mu(\omega_1)\\
    &\times[q_{\bm k}^{\mu},
    \rho^{(p)}(\bm k,\omega-\omega_1)]_{mn}.
    \end{aligned}
    \label{eq:thermal-density-matrix-recursion-main}
\end{equation}
The positive infinitesimal $\eta=0^+$ ensures causality for the retarded response. A finite phenomenological broadening $\Gamma$ is introduced separately below.
The zeroth-order term is the equilibrium Fermi distribution, $\rho_{mn}^{(0)}(\bm k,\omega)=2\pi\delta(\omega)f_{\bm k n}\delta_{mn}$.
Setting $p=0$ in Eq.~\eqref{eq:thermal-density-matrix-recursion-main} gives the first-order response to the thermal field.
Its diagonal element describes the change in band populations:
\begin{equation}
    \rho_{nn}^{(1)}(\bm k,\omega)
    =-\frac{E_T^\nu(\omega)\widetilde{\mathcal E}_{\bm k n}}
    {\hbar\omega+i\eta}
    i\partial_{k_\nu}f_{\bm k n}.
    \label{eq:rho-diagonal}
\end{equation}
The off-diagonal element describes thermally induced interband coherence:
\begin{equation}
    \begin{aligned}
    \rho_{mn}^{(1)}(\bm k,\omega)
    &=\frac{E_T^\nu(\omega)}
    {2(\hbar\omega-\mathcal E_{\bm k,mn}+i\eta)}
    (\widetilde{\mathcal E}_{\bm k m}
    +\widetilde{\mathcal E}_{\bm k n})\\
    &\quad\times
    \mathcal A_{mn}^{\nu}f_{\bm k,mn},\qquad m\ne n.
    \end{aligned}
    \label{eq:rho-offdiagonal}
\end{equation}
Here, $\widetilde{\mathcal E}_{\bm k n}=\mathcal E_{\bm k n}-\mu$ is the band energy relative to the chemical potential, $\mathcal E_{\bm k,mn}=\mathcal E_{\bm k m}-\mathcal E_{\bm k n}$ is the interband energy difference, and $f_{\bm k,mn}=f_{\bm k m}-f_{\bm k n}$ is the occupation difference.

\subsection{Thermoelectric coefficients and intraband--interband decomposition}

The velocity operator is $v_{\bm k}^{\mu}=[r_{\bm k}^{\mu},h_{\mathrm{MF}}]/(i\hbar)$. Its matrix elements in the interacting basis are $v_{\bm k,mn}^{\mu}=\hbar^{-1} \partial_{k_\mu}\mathcal E_{\bm k m}\delta_{mn} +i\mathcal E_{\bm k,mn}\mathcal A_{mn}^{\mu}/\hbar$.
With $-e$ the electron charge and $e>0$, the first-order density matrix gives the local Kubo current $j_e^{\mathrm{Kubo},\mu}=-e\intk\sum_{mn}\rho_{mn}^{(1)}v_{nm}^{\mu}$. We define the Kubo thermoelectric tensor by $j_e^{\mathrm{Kubo},\mu}=L^{\mathrm{Kubo},\mu\nu}E_T^\nu$.
The transport thermoelectric tensor is defined by $j_e^{\mathrm{tr},\mu}=\alpha^{\mu\nu}(-\partial_\nu T)$.
The local Kubo response contains both the transport current and an orbital-magnetization current associated with equilibrium circulation.
These circulating currents do not transfer charge through the sample, so their field-induced contribution must be removed to obtain the transport response.
With the present tensor and thermal-field conventions, this transport thermoelectric tensor $\alpha^{\mu\nu}$ is constructed as
\begin{equation}
    \alpha^{\mu\nu}
    =\frac{1}{T}\left(
    \operatorname{Re}L_{\mathrm{dc}}^{\mathrm{Kubo},\mu\nu}
    +M_{\mathrm{orb}}^{\mu\nu}\right),
\label{eq:alpha-from-kubo-and-magnetization-main}
\end{equation}
where $M_{\mathrm{orb}}^{\mu\nu}$ is the orbital-magnetization tensor, which is antisymmetric under exchange of the spatial indices, $M_{\mathrm{orb}}^{\mu\nu}=-M_{\mathrm{orb}}^{\nu\mu}$.
The diagonal and off-diagonal parts of $\rho^{(1)}$ give the intraband and interband responses. Separating the latter into parts symmetric and antisymmetric under $\mu\leftrightarrow\nu$ yields three thermoelectric channels:
\begin{equation}
    \alpha^{\mu\nu}
    =\alpha_{\mathrm{intra}}^{\mu\nu}
    +\alpha_{\mathrm{inter,s}}^{\mu\nu}
    +\alpha_{\mathrm{inter,a}}^{\mu\nu}.
\label{eq:three-alpha-channels-main}
\end{equation}
The density-matrix solution and the intraband--interband separation are derived in Appendix~\ref{app:thermoelectric-response}.

The corresponding frequency-dependent intraband coefficient is
\begin{equation}
    \begin{aligned}
    \alpha_{\mathrm{intra}}^{\mu\nu}(\omega)
    =-&\frac{e\eta}{T[(\hbar\omega)^2+\eta^2]}
    \intk\sum_n f_{\bm k n}\\
    &\times\partial_{k_\nu}
    \bigl(\widetilde{\mathcal E}_{\bm k n}v_{\bm k n}^{\mu}\bigr).
    \end{aligned}
    \label{eq:intraband-dynamical-alpha}
\end{equation}
Here, $v_{\bm k n}^{\mu}=\hbar^{-1}\partial_{k_\mu}\mathcal E_{\bm k n}$ is the group velocity, and $\partial_{k_\nu}(\widetilde{\mathcal E}_{\bm k n}v_{\bm k n}^{\mu})$ is the momentum derivative of the band-diagonal heat-current matrix element.
The intraband coefficient $\alpha_{\mathrm{intra}}^{\mu\nu}$ is symmetric in its spatial indices, depends only on the renormalized band dispersion, and gives the real Drude response.

For the interband response, the Kubo coefficient can be written in terms of the matrix elements of the quantum geometric tensor,
\begin{equation}
    \begin{aligned}
    L_{\mathrm{inter}}^{\mathrm{Kubo},\mu\nu}(\omega)
    =-&\frac{ie}{2\hbar}\intk\sum_{m\ne n}
    \frac{\mathcal E_{\bm k,nm}}
    {\hbar\omega-\mathcal E_{\bm k,mn}+i\eta}\\
    &\times(\widetilde{\mathcal E}_{\bm k m}
    +\widetilde{\mathcal E}_{\bm k n})
    f_{\bm k,mn}\mathcal Q_{mn}^{\nu\mu}.
    \end{aligned}
    \label{eq:interband-kubo-qgt-main}
\end{equation}
Here, $\mathcal Q_{mn}^{\mu\nu}=\mathcal A_{mn}^{\mu}\mathcal A_{nm}^{\nu}$ for $m\ne n$, with the Berry connection defined above.
It decomposes as $\mathcal Q_{mn}^{\mu\nu}=g_{mn}^{\mu\nu}-i\Omega_{mn}^{\mu\nu}/2$, where $g_{mn}^{\mu\nu}=\tfrac12(\mathcal A_{mn}^{\mu}\mathcal A_{nm}^{\nu}+\mathcal A_{mn}^{\nu}\mathcal A_{nm}^{\mu})$ is the quantum-metric matrix element and $\Omega_{mn}^{\mu\nu}=i(\mathcal A_{mn}^{\mu}\mathcal A_{nm}^{\nu}-\mathcal A_{mn}^{\nu}\mathcal A_{nm}^{\mu})$ is the Berry-curvature matrix element. The band-resolved Berry curvature is $\Omega_n^{\mu\nu}=\sum_{m\ne n}\Omega_{nm}^{\mu\nu}$.
The quantum metric is symmetric under exchange of spatial indices, $g_{mn}^{\mu\nu}=g_{mn}^{\nu\mu}$, whereas the Berry curvature is antisymmetric, $\Omega_{mn}^{\mu\nu}=-\Omega_{mn}^{\nu\mu}$.
The symmetric part of the interband thermoelectric coefficient is therefore governed by the quantum metric,
\begin{equation}
    \begin{aligned}
    \alpha_{\mathrm{inter,s}}^{\mu\nu}(\omega)
    =-&\frac{\pi e}{2\hbar T}\intk\sum_{m\ne n}
    \mathcal E_{\bm k,nm}\\
    &\times\left[\delta(\hbar\omega-\mathcal E_{\bm k,mn})
    +\delta(\hbar\omega-\mathcal E_{\bm k,nm})\right]\\
    &\times
    (\widetilde{\mathcal E}_{\bm k m}
    +\widetilde{\mathcal E}_{\bm k n})
    g_{mn}^{\mu\nu}f_{\bm k m}.
    \end{aligned}
    \label{eq:interband-symmetric-alpha-main}
\end{equation}
For nondegenerate bands, the interband resonance condition enforced by the delta functions cannot be met in the clean dc limit ($\omega\to0$), so this symmetric contribution vanishes. Finite broadening can produce a dc response, as discussed in \Sec{sec:scattering-symmetry}.

For the antisymmetric channel, accounting for orbital magnetization cancels the orbital-moment term in the Kubo coefficient, yielding the Berry-curvature thermoelectric response below.
The resulting coefficient is
\begin{equation}
    \alpha_{\mathrm{inter,a}}^{\mu\nu}
    =\frac{e k_{\mathrm B}}{\hbar}\intk
    \sum_{m\ne n}\Omega_{mn}^{\mu\nu}S(f_{\bm k m}),
    \label{eq:entropy-weighted-alpha}
\end{equation}
where $S(f)=-f\ln f-(1-f)\ln(1-f)$ is the entropy density.
Equation~\eqref{eq:entropy-weighted-alpha} has the same form as the noninteracting semiclassical result \cite{Xiao2006}, but all quantities now belong to the equilibrium self-consistent Hamiltonian. Interactions modify both the energies entering $S(f)$ and the Bloch-state geometry encoded in $\Omega_{mn}^{\mu\nu}$.

\section{Bloch Geometry from Interactions}
\label{sec:interaction-geometry}

Equation~\eqref{eq:entropy-weighted-alpha} expresses the anomalous response in terms of the reconstructed bands.
To identify the contribution of the interaction connection to the Berry-curvature response, we express the interacting Berry connection in the noninteracting basis.
Let $A^\mu$ denote the Berry connection in the noninteracting basis. The unitary matrix $W(\bm k)$ relates the noninteracting Bloch basis to the self-consistent interacting eigenbasis through $\ket{n(\bm k)}=\sum_a W_{an}(\bm k)\ket{a(\bm k)}$.
Transforming the interacting-basis connection back to the noninteracting basis gives
\begin{equation}
\widetilde{\mathcal A}^{\mu}
\equiv W\mathcal A^{\mu}W^\dagger=A^\mu+\mathcal I^\mu.
    \label{eq:transformed-mean-field-connection}
\end{equation}
The interaction connection is
\begin{equation}
\mathcal I^\mu=i(\partial_{k_\mu}W)W^\dagger.
\label{eq:interaction-induced-connection}
\end{equation}
In this reference, the matrix $W(\bm k)$ describes the eigenstate reconstruction induced by the HF self-energy, while the interaction connection $\mathcal I^\mu$ explicitly gives the additional contribution of this reconstruction to the Berry connection. The connection decomposition therefore makes explicit how eigenstate reconstruction enters the geometric response. Interactions also affect the response through changes in the energy spectrum and the factors $S(f)$ entering the response.
Here and below in this representation, $\rho(\bm k)=WfW^\dagger$ denotes the equilibrium density matrix in the noninteracting basis, with $f(\bm k)=\operatorname{diag}(f_{\bm k n})$; momentum arguments are suppressed when unambiguous.
The function $S(f)$ is represented in the reference basis by the matrix function $S(\rho)=-\rho\ln\rho-(I-\rho)\ln(I-\rho)=WS(f)W^\dagger$.
Rewriting the band-index sum in Eq.~\eqref{eq:entropy-weighted-alpha} as a matrix trace and expanding $\widetilde{\mathcal A}^{\mu}=A^\mu+\mathcal I^\mu$ yields
\begin{equation}
\alpha_{\mathrm{inter,a}}^{\mu\nu}
=\frac{e k_{\mathrm B}}{\hbar}\intk
\operatorname{Tr}\!\left[S(\rho)\mathcal C_{\mathrm{tot}}^{\mu\nu}\right].
\label{eq:interaction-resolved-alpha}
\end{equation}
Here $\mathcal C_{\mathrm{tot}}^{\mu\nu}=\mathcal C_{\mathrm{bare}}^{\mu\nu}+\mathcal C_{\mathrm{coh}}^{\mu\nu}+\mathcal C_{\mathrm{int}}^{\mu\nu}$.
The bare geometric contribution is
\begin{equation}
\mathcal C_{\mathrm{bare}}^{\mu\nu}
=i[A^\mu,A^\nu].
\label{eq:bare-curvature-contribution}
\end{equation}
The generator $\mathcal C_{\mathrm{bare}}^{\mu\nu}$ is constructed from the noninteracting connection but is weighted by the interacting entropy matrix $S(\rho)$, which contains both occupations and coherences in that basis. ``Bare'' therefore refers to the geometric generator; its transport contribution generally differs from the response of the noninteracting system.

The coherence contribution is
\begin{equation}
\mathcal C_{\mathrm{coh}}^{\mu\nu}
=i[A^\mu,\mathcal I^\nu]+i[\mathcal I^\mu,A^\nu].
\label{eq:mixed-curvature-contribution}
\end{equation}
The coherence term couples the bare connection $A^\mu$ to the interaction connection $\mathcal I^\mu$.

Finally, the term built solely from the interaction connection is
\begin{equation}
\mathcal C_{\mathrm{int}}^{\mu\nu}
=i[\mathcal I^\mu,\mathcal I^\nu].
\label{eq:interaction-curvature-contribution}
\end{equation}
Within this trace expression, $\mathcal C_{\mathrm{tot}}^{\mu\nu}=i[\widetilde{\mathcal A}^{\mu},\widetilde{\mathcal A}^{\nu}]$ serves as the curvature generator.

In the noninteracting limit, choosing identical bases gives $W(\bm k)=I$ and $\mathcal I^\mu=0$. The coherence and interaction-induced terms vanish, leaving the standard Berry-curvature thermoelectric response from $\mathcal C_{\mathrm{bare}}^{\mu\nu}$.
If $\mathcal C_{\mathrm{bare}}^{\mu\nu}=0$, the anomalous response comes from the coherence and interaction-connection terms. If $A^\mu=0$ as well, only $\mathcal C_{\mathrm{int}}^{\mu\nu}$ contributes.

The interpretation of this decomposition depends on the reference basis. We retain an interaction-independent eigenbasis of $h_0$, in which every off-diagonal matrix element of $h_{\mathrm{MF}}$ comes from the HF self-energy. A momentum-dependent change of reference transforms both $h_0$ and the self-energy and can redistribute contributions between the reference and interaction connections. The bare, coherence, and interaction terms must then be interpreted relative to the new reference. Our fixed choice keeps the distinction between noninteracting geometry and the additional connection generated by interactions explicit.

\section{Broadening and symmetry classification of thermoelectric channels}
\label{sec:scattering-symmetry}


\subsection{Finite-broadening dc response}

We use the relaxation-time approximation with a momentum-independent scalar energy broadening $\Gamma=\hbar/\tau$, where $\tau$ is the relaxation time.
We apply the same scalar broadening to the diagonal and off-diagonal components of the density-matrix response.
Within this approximation, the density matrix obeys $i\hbar\dot\rho=[h_{\mathrm{MF}}+h_T,\rho]-i\Gamma(\rho-\rho^{(0)})$.
The intraband coefficient follows from \Eq{eq:intraband-dynamical-alpha} as
\begin{equation}
 \alpha_{\mathrm{intra}}^{\mu\nu}
 =-\frac{e}{T\Gamma}\intk\sum_n f_{\bm k n}
 \partial_{k_\nu}(\widetilde{\mathcal E}_{\bm k n}v_{\bm k n}^{\mu}),
 \label{eq:finite-Gamma-Drude}
\end{equation}
and therefore scales as $\Gamma^{-1}$.

\begin{table}[!b]
\caption{Band quantities, tensor symmetry, and leading broadening dependence of the three thermoelectric contributions within the relaxation-time approximation.
Momentum labels are suppressed in the band quantities; $\widetilde{\mathcal E}_n v_n^\mu$ is the band-diagonal heat-current matrix element, with $\widetilde{\mathcal E}_n=\mathcal E_n-\mu$.
The interband scalings apply away from band degeneracies when $\Gamma$ is much smaller than the relevant interband separations.}
\label{tab:channel-classification-main}
\renewcommand{\arraystretch}{1.2}
\setlength{\tabcolsep}{3pt}
\begin{ruledtabular}
\begin{tabular}{lccc}
Channel & Tensor & Band quantity & Scaling\\
\hline
$\alpha_{\mathrm{intra}}$ & symmetric & $\partial_{k_\nu}(\widetilde{\mathcal E}_n v_n^\mu)$ & $\Gamma^{-1}$\\
$\alpha_{\mathrm{inter,s}}$ & symmetric & $g^{\mu\nu}$ & $\Gamma$\\
$\alpha_{\mathrm{inter,a}}$ & antisymmetric & $\Omega^{\mu\nu}$ & $\Gamma^0$
\end{tabular}
\end{ruledtabular}
\end{table}
The same broadening produces a dissipative symmetric interband coefficient.
Away from band degeneracies and for $\Gamma\ll|\mathcal E_{\bm k,mn}|$,
\begin{equation}
    \alpha_{\mathrm{inter,s}}^{\mu\nu}
    =-\frac{e\Gamma}{T\hbar}\intk
    \sum_{m\ne n}G_{mn}^{\mu\nu}f_{\bm k m},
    \label{eq:finite-Gamma-metric}
\end{equation}
where
\begin{equation}
    G_{mn}^{\mu\nu}
    =-\frac{\widetilde{\mathcal E}_{\bm k m}
    +\widetilde{\mathcal E}_{\bm k n}}
    {\widetilde{\mathcal E}_{\bm k m}
    -\widetilde{\mathcal E}_{\bm k n}}
    g_{mn}^{\mu\nu}.
    \label{eq:thermal-connection-polarizability}
\end{equation}
The matrix elements $G_{mn}^{\mu\nu}$ of the thermal Berry-connection polarizability also contribute to the intrinsic nonlinear thermoelectric response \cite{Wang2022b,Li2024}.
This contribution is linear in $\Gamma$ when the relevant interband separations are much larger than $\Gamma$. A related separation between band-population and geometric interband contributions appears in electrical conductivity \cite{Mitscherling2020}.

The antisymmetric Berry-curvature coefficient $\alpha_{\mathrm{inter,a}}$ describes the intrinsic anomalous response.
For $\Gamma\ll|\mathcal E_{\bm k,mn}|$, the Berry-curvature contribution retains its clean-limit value to leading order in the broadening, giving $\alpha_{\mathrm{inter,a}}\propto\Gamma^0$.

\subsection{General symmetry constraints}

We now examine the spatial and time-reversal symmetry conditions under which the Drude, quantum-metric, and Berry-curvature channels contribute to transverse transport.

\begin{table*}[!t]
\caption{Symmetry selection rules for transverse two-dimensional charge thermoelectric coefficients.
A scalar symmetry-preserving broadening and zero external magnetic field are assumed.
A check mark means that the symmetry allows the contribution, not that it must be nonzero.
Here, $\mathcal M_x:(x,y,z)\mapsto(-x,y,z)$ denotes reflection in the $yz$ plane.}
\label{tab:general-symmetry-main}
\footnotesize
\setlength{\tabcolsep}{2pt}
\renewcommand{\arraystretch}{1.25}
\begin{tabularx}{\textwidth}{
>{\raggedright\arraybackslash}p{0.19\textwidth}|
*{9}{>{\centering\arraybackslash}X}}
\hline\hline
\rule{0pt}{3.0ex}\multirow[c]{2}{*}{Contribution}
&\multirow[c]{2}{*}{$\mathcal P$}
&\multirow[c]{2}{*}{$\mathcal T$}
&\multirow[c]{2}{*}{$\mathcal{PT}$}
&\multirow[c]{2}{*}{$C_{2z}$}
&$C_{nz}$
&\multirow[c]{2}{*}{$\mathcal M_x$}
&\multirow[c]{2}{*}{$C_{2z}\mathcal T$}
&$C_{nz}\mathcal T$
&\multirow[c]{2}{*}{$\mathcal M_x\mathcal T$}\\
\rule[-1.0ex]{0pt}{2.7ex}
&&&&&$(n>2)$&&&$(n>2)$&\\
\hline
$\alpha_{\mathrm{intra}}^{xy}$
&$\checkmark$&$\checkmark$&$\checkmark$
&$\checkmark$&$\times$&$\times$
&$\checkmark$&$\times$&$\times$\\
$\alpha_{\mathrm{inter,s}}^{xy}$
&$\checkmark$&$\checkmark$&$\checkmark$
&$\checkmark$&$\times$&$\times$
&$\checkmark$&$\times$&$\times$\\
$\alpha_{\mathrm{inter,a}}^{xy}$
&$\checkmark$&$\times$&$\times$
&$\checkmark$&$\checkmark$&$\times$
&$\times$&$\times$&$\checkmark$\\
\hline\hline
\end{tabularx}
\end{table*}

Decomposing the total thermoelectric tensor into its symmetric and antisymmetric components yields
\begin{equation}
 \alpha_{\mathrm{s}}=\begin{pmatrix}
\alpha^{xx}&\alpha_{\mathrm{s}}^{xy}\\ \alpha_{\mathrm{s}}^{xy}&\alpha^{yy}
\end{pmatrix},\quad
\alpha_{\mathrm{a}}=\begin{pmatrix}
0&\alpha_{\mathrm{a}}^{xy}\\ -\alpha_{\mathrm{a}}^{xy}&0
\end{pmatrix}.
\end{equation}
The symmetric part is the sum of the Drude and quantum-metric contributions, $\alpha_{\mathrm{s}}=\alpha_{\mathrm{intra}}+\alpha_{\mathrm{inter,s}}$. The antisymmetric part is the Berry-curvature contribution, $\alpha_{\mathrm{a}}=\alpha_{\mathrm{inter,a}}$.

If the state is invariant under an in-plane rotation $R(\vartheta)$, its thermoelectric tensor satisfies $\alpha=R(\vartheta)\alpha R^T(\vartheta)$.
For rotations of order $n>2$, these constraints require longitudinal isotropy and a vanishing symmetric transverse response:
\begin{equation}
 \alpha^{xx}=\alpha^{yy},\qquad \alpha_{\mathrm{s}}^{xy}=0.
 \label{eq:rotation-symmetric-constraints}
\end{equation}
Both conditions hold separately for the Drude and quantum-metric channels.
A finite symmetric transverse response therefore requires breaking the rotational symmetry that enforces isotropy of the charge response. Such symmetry breaking can arise from self-consistent electronic order even when the bare lattice has higher rotational symmetry.
Proper in-plane rotations do not forbid the antisymmetric Berry-curvature response.
Reflection in the $yz$ plane is represented in the $xy$ plane by $\mathcal M_x=\operatorname{diag}(-1,1)$. Mirror invariance requires $\alpha=\mathcal M_x\alpha\mathcal M_x^T$, which gives $\alpha^{xy}=\alpha^{yx}=0$.

Spatial inversion ($\mathcal P$) and time reversal ($\mathcal T$) impose different constraints on the geometric matrix elements.
As detailed in Appendix~\ref{app:symmetry-details}, the Berry curvature $\Omega$ is $\mathcal P$-even and $\mathcal T$-odd, whereas the quantum metric $g$ is even under both $\mathcal P$ and $\mathcal T$.
Spatial inversion alone does not forbid any of the three transport channels. Time-reversal symmetry forbids the antisymmetric Berry-curvature response but permits the Drude and quantum-metric contributions.
Table~\ref{tab:general-symmetry-main} summarizes these selection rules, including those imposed by combined spatial and time-reversal operations.

\section{Phase Diagram and Transverse Thermoelectric Responses}
\label{sec:triangular-results}

We first examine how complex spin-flip mean fields modify the Berry curvature and quantum metric in a two-band model and enter its thermoelectric response.
We then apply this analysis to the triangular-lattice extended Hubbard model, focusing on transverse transport and rotational symmetry.

\subsection{Complex spin-flip order and two-band geometry}
\label{sec:two-band-chirality}

We consider a single-orbital model with two spin components coupled by complex off-diagonal mean fields.
To examine the band geometry, we take a constant diagonal splitting and omit the common scalar dispersion, which does not affect the eigenstates. The mean-field Hamiltonian is then
\begin{equation}
 h_{\mathrm{MF}}(\bm k)=
 \begin{pmatrix}
 m_0&\mathcal D^*(\bm k)\\
 \mathcal D(\bm k)&-m_0
 \end{pmatrix},
 \label{eq:generic-constant-diagonal-model}
\end{equation}
where $2m_0$ is the constant diagonal splitting and $\mathcal D(\bm k)$ is the complex off-diagonal mean field.
For this Hamiltonian, the Berry curvature is \cite{Xiao2010,Sticlet2012} (see Appendix~\ref{app:mean-field-quantum-metric} for the derivation)
\begin{equation}
 \Omega_\pm^{xy}(\bm k)=\mp
 \frac{m_0\operatorname{Im}[\partial_{k_x}\mathcal D^*\,
                             \partial_{k_y}\mathcal D]}
 {2[m_0^2+|\mathcal D(\bm k)|^2]^{3/2}}.
 \label{eq:generic-constant-mass-curvature}
\end{equation}
For this constant-splitting model, nonzero local Berry curvature requires both a finite diagonal splitting and a nonzero imaginary part of the mean-field gradient product.
In a translationally invariant single-orbital system where inversion acts trivially on spin, inversion-symmetric onsite and nearest-neighbor mean fields admit the parametrization
\begin{equation}
 \mathcal D(\bm k)=\mathcal D_0
 +2\Lambda\sum_\ell\mathcal C_\ell\cos(\bm k\cdot\bm\eta_\ell).
 \label{eq:generic-cosine-order}
\end{equation}
Here, $\mathcal D_0$ is a constant onsite field, $\Lambda$ is a real energy scale, and $\mathcal C_\ell$ is the complex dimensionless bond amplitude along $\bm\eta_\ell$.
The imaginary part entering the Berry curvature is a sum over bond pairs (Appendix~\ref{app:mean-field-quantum-metric}):
\begin{equation}
 \begin{aligned}
 &\operatorname{Im}[\partial_{k_x}\mathcal D^*\,\partial_{k_y}\mathcal D]
 =4\Lambda^2\sum_{\ell<\ell'}(\bm\eta_\ell\times\bm\eta_{\ell'})_z\\
 &\qquad\times\operatorname{Im}(\mathcal C_\ell^*\mathcal C_{\ell'})
 \sin(\bm k\cdot\bm\eta_\ell)\sin(\bm k\cdot\bm\eta_{\ell'}).
 \end{aligned}
 \label{eq:generic-pairwise-chirality}
\end{equation}
Here $\operatorname{Im}(\mathcal C_\ell^*\mathcal C_{\ell'})=|\mathcal C_\ell||\mathcal C_{\ell'}|\sin(\phi_{\ell'}-\phi_\ell)$, with $\phi_\ell=\arg\mathcal C_\ell$. A relative phase different from $0$ or $\pi$ on at least one noncollinear bond pair is necessary for a nonzero sum, but cancellations can still make it vanish. Such phase-dependent fields provide a route to interaction-generated Berry curvature \cite{ZhangTewariYakovenkoDasSarma2008,KotetesVarelogiannis2010}.
Different bonds enter $\mathcal D(\bm k)$ with different momentum-dependent weights. Nontrivial relative bond phases can therefore give this spin-flip field a momentum-dependent phase, changing the relative phase of the two spin components of a Bloch eigenstate. This variation allows finite diagonal quantum-metric components even for $m_0=0$, where the Berry curvature vanishes away from band touchings (Appendix~\ref{app:mean-field-quantum-metric}).
A finite local $g^{xy}$ does not by itself imply transverse transport, which also depends on the symmetry of the charge response.

\begin{figure*}[!t]
\centering
\begin{minipage}[c]{0.49\textwidth}
\centering
\includegraphics[width=\linewidth]{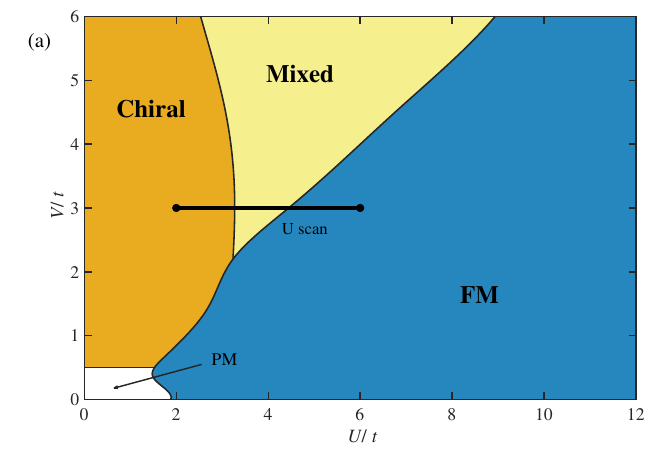}
\end{minipage}\hfill
\begin{minipage}[c]{0.49\textwidth}
\centering
\includegraphics[width=\linewidth]{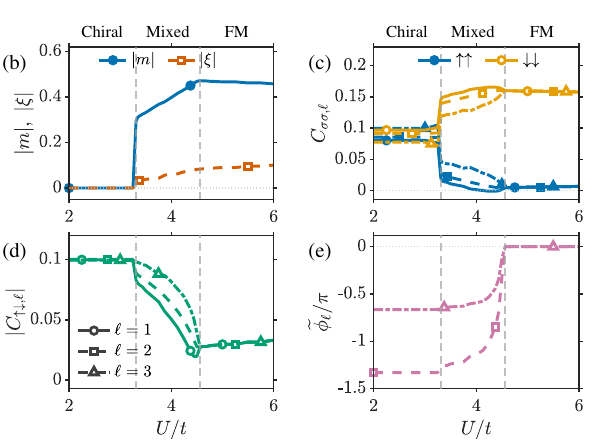}
\end{minipage}
\caption{Classification of self-consistent states and their order parameters at $n=1.5$ and $k_{\mathrm B}T/t=0.01$, without bare spin--orbit coupling or a Zeeman field.
\textbf{(a)} The $U$--$V$ map classified in a fixed spin basis by the $z$-component magnetization $m=n_\uparrow-n_\downarrow$ and chiral bond order: the chiral region has chiral bond order with $m=0$, the ferromagnetic (FM) region has finite $m$ without chiral bond order, the mixed region has both, and the remaining weak-coupling region is labeled paramagnetic metal (PM). The classification criteria are given in Appendix~\ref*{app:model-checks-materials}.
The curves in (a) schematically indicate the phase boundaries of the HF solutions. The remaining panels show the order parameters of the chiral, mixed, and FM states at fixed $V=3t$.
\textbf{(b)} The magnitudes $|m|$ and $|\xi|$ of the $z$-component magnetization $m$ and onsite spin coherence.
\textbf{(c)} The signed spin-conserving bond fields $C_{\uparrow\uparrow,\ell}$ and $C_{\downarrow\downarrow,\ell}$ for $\ell=1,2,3$; filled markers denote spin-up fields and open markers denote spin-down fields.
\textbf{(d)} The spin-flip bond magnitudes $|C_{\uparrow\downarrow,\ell}|$.
\textbf{(e)} The continuously unwrapped relative phases $\widetilde{\phi}_\ell/\pi$ with respect to $C_1$, for $\ell=2,3$ and $C_\ell=C_{\uparrow\downarrow,\ell}$.
Bond directions use the same line styles and marker shapes in \textbf{(c)}--\textbf{(e)}. The vertical dashed lines in Figs.~1--3 schematically indicate the phase boundaries between the chiral, mixed, and FM states.}
\label{fig:preliminary-order}
\end{figure*}

\subsection{Triangular-lattice realization and self-consistent states}
\label{sec:triangular-realization}

We now consider the extended Hubbard model on a one-site triangular Bravais lattice:
\begin{equation}
    \begin{aligned}
    H={}&-t\sum_{\langle ij\rangle,\sigma}
    (c_{i\sigma}^\dagger c_{j\sigma}+\mathrm{H.c.})\\
    &+U\sum_i n_{i\uparrow}n_{i\downarrow}
    +V\sum_i\sum_{\bm\delta\in\mathcal N}n_i n_{i+\bm\delta}.
    \end{aligned}
    \label{eq:triangular-model-main}
\end{equation}
Here, $i$ and $j$ label lattice sites, $\sigma=\uparrow,\downarrow$ denotes the spin index, and $\langle ij\rangle$ represents an unordered nearest-neighbor bond counted once.
The operator $c_{i\sigma}^{\dagger}$ creates a fermion with spin $\sigma$ at site $i$, $n_{i\sigma}=c_{i\sigma}^{\dagger}c_{i\sigma}$, and $n_i=n_{i\uparrow}+n_{i\downarrow}$. Here, H.c.\ denotes the Hermitian conjugate.
Here, $t$ is the real nearest-neighbor hopping amplitude, $U$ the onsite repulsion, and $V$ the nearest-neighbor density--density repulsion.
The three independent nearest-neighbor vectors are $\bm\eta_1=a(1,0)$, $\bm\eta_2=a(-1/2,-\sqrt3/2)$, and $\bm\eta_3=a(-1/2,\sqrt3/2)$, where $a$ is the nearest-neighbor distance (the lattice constant) and $\ell=1,2,3$ labels the distinct bond directions. These representatives are separated by $120^\circ$ and satisfy $\bm\eta_1+\bm\eta_2+\bm\eta_3=0$.
The interaction sum employs $\mathcal N=\{\pm\bm\eta_1,\pm\bm\eta_2, \pm\bm\eta_3\}$.
Using the momentum-space transformation $c_{\bm k\sigma}=N^{-1/2}\sum_i e^{-i\bm k\cdot\bm R_i}c_{i\sigma}$, the bare noninteracting dispersion is given by $\epsilon_{\bm k}^{(0)}=-2t\sum_{\ell=1}^{3} \cos(\bm k\cdot\bm\eta_\ell)$.

The onsite and nearest-neighbor interactions generate the onsite and bond mean fields introduced above. Appendix~\ref{app:triangular-hf-symmetry} derives their HF contributions and self-consistency equations. We compare the symmetry of the resulting transport response with the sixfold symmetry of the bare lattice.
We use an HF decoupling that preserves translation and spatial inversion symmetries.
The HF Hamiltonian in \Eq{eq:general-HMF} contains the constant term $E_{\mathrm{const}}$ and the one-body term $\sum_{\bm k}\Psi_{\bm k}^{\dagger}h_{\mathrm{MF}}(\bm k)\Psi_{\bm k}$. For the spinor $\Psi_{\bm k}=(c_{\bm k\uparrow},c_{\bm k\downarrow})^T$, the corresponding single-particle matrix is
\begin{equation}
    \begin{gathered}
    h_{\mathrm{MF}}(\bm k)=\\[2pt]
    \begin{pmatrix}
    \epsilon_{\bm k}-\dfrac{U}{2}m
    -2V\Delta_{\uparrow\uparrow}(\bm k)
    &-U\xi^*-2V\Delta_{\uparrow\downarrow}^*(\bm k)\\[3pt]
    -U\xi-2V\Delta_{\uparrow\downarrow}(\bm k)
    &\epsilon_{\bm k}+\dfrac{U}{2}m
    -2V\Delta_{\downarrow\downarrow}(\bm k)
    \end{pmatrix}.\end{gathered}
    \label{eq:triangular-HMF-main}
\end{equation}
Here, $\epsilon_{\bm k}=\epsilon_{\bm k}^{(0)}+Un/2+12Vn$ contains the bare dispersion and the uniform Hartree shift.
For this one-site Bravais lattice, the number of sites is $N=N_{\mathrm c}$.
The occupation per site for spin $\sigma$ is $n_\sigma=N^{-1}\sum_{\bm k}\langle c_{\bm k\sigma}^{\dagger} c_{\bm k\sigma}\rangle$.
We define the filling per site as $n=n_\uparrow+n_\downarrow$ and the dimensionless $z$-component magnetization per site as $m=n_\uparrow-n_\downarrow$.
The onsite Fock term also includes the onsite spin coherence $\xi=N^{-1}\sum_{\bm k}\langle c_{\bm k\uparrow}^{\dagger}c_{\bm k\downarrow}\rangle$.

For spin indices $\sigma,\sigma'$ and bond direction $\ell$, we define the nearest-neighbor bond field as
\begin{equation}
C_{\sigma\sigma',\ell}
=\frac{1}{N}\sum_{\bm k}
\cos(\bm k\cdot\bm\eta_\ell)
\left\langle c_{\bm k\sigma}^{\dagger}
c_{\bm k\sigma'}\right\rangle.
\label{eq:triangular-directional-bond-order-main}
\end{equation}
The corresponding momentum-space form factor is
\begin{equation}
\Delta_{\sigma\sigma'}(\bm k)
=2\sum_{\ell=1}^3C_{\sigma\sigma',\ell}
\cos(\bm k\cdot\bm\eta_\ell).
\label{eq:triangular-bond-field-main}
\end{equation}
The $\sigma=\sigma'$ channel describes spin-conserving bond fields $C_{\sigma\sigma,\ell}$, whereas the $\sigma\ne\sigma'$ channel describes spin-flip bond fields $C_{\uparrow\downarrow,\ell}$ and their complex conjugates.
Comparison with Eq.~\eqref{eq:generic-cosine-order} gives $\mathcal D_0=-U\xi$, $\Lambda=-2V$, and $\mathcal C_\ell=C_{\uparrow\downarrow,\ell}$.

The spin-flip bond fields $C_{\uparrow\downarrow,\ell}$ can form chiral bond order, with a representative configuration given by
\begin{equation}
 C_{\uparrow\downarrow,\ell}
 =|C_{\uparrow\downarrow,\ell}|
 e^{i[\phi+2\pi(\ell-1)/3]},\qquad \ell=1,2,3,
 \label{eq:triangular-chiral-bond-pattern}
\end{equation}
where $\phi$ is a global phase and the bond amplitudes may differ between directions.
Through \Eq{eq:triangular-HMF-main}, these complex bond fields can modify both the Berry curvature and quantum metric.

Table~\ref{tab:general-symmetry-main} shows that sixfold symmetry of the charge response enforces equal longitudinal components and forbids symmetric transverse transport. Nontrivial relative phases in the spin-flip bond fields $C_{\uparrow\downarrow,\ell}$ break pure lattice $C_6$ symmetry while preserving the twofold rotation associated with spatial inversion. Equal-amplitude chiral bond order can nevertheless retain $C_6$ symmetry of the charge response through a compensating spin rotation (Appendix~\ref{app:combined-lattice-spin}).

\begin{figure*}[!t]
\centering
\includegraphics[width=\textwidth]{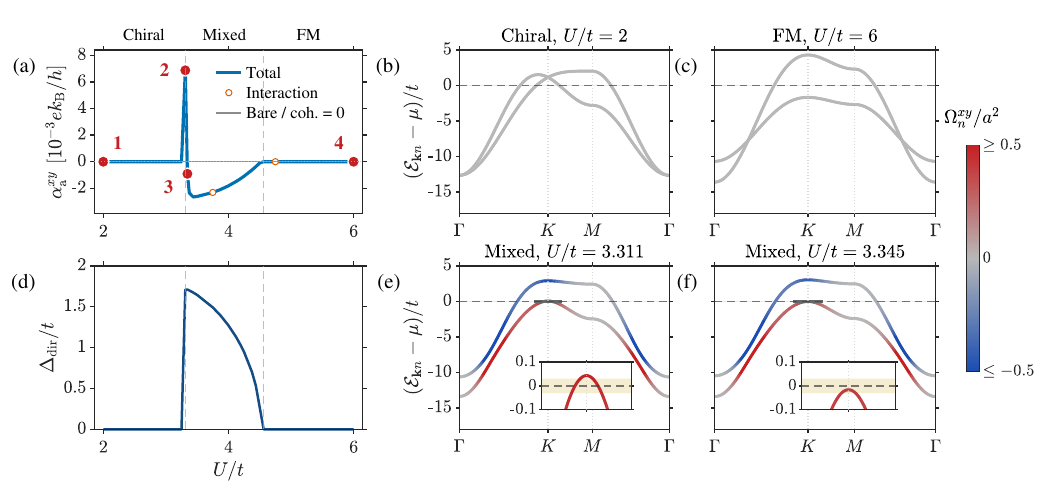}
\caption{Anomalous thermoelectric response and band reconstruction along the fixed-$V/t=3$ interaction scan at $n=1.5$ and $k_{\mathrm B}T/t=0.01$.
\textbf{(a)} The anomalous transverse coefficient $\alpha_{\mathrm a}^{xy}$ and its bare, coherence, and interaction-induced contributions. The total response coincides with the interaction-induced contribution. Red markers identify the four selected states.
\textbf{(b)} Band spectrum of the chiral state (point 1).
\textbf{(c)} Band spectrum of the FM state (point 4).
\textbf{(d)} The direct band gap $\Delta_{\mathrm{dir}}/t$ across the Brillouin zone.
\textbf{(e)} Band spectrum of the mixed state before the response sign reversal (point 2).
\textbf{(f)} Band spectrum of the mixed state after the response sign reversal (point 3).
The band spectra follow $\Gamma$--$K$--$M$--$\Gamma$, are plotted as $(\mathcal E_{\bm k n}-\mu)/t$, and are colored by the Berry curvature $\Omega_n^{xy}/a^2$ on a shared scale. The insets enlarge the vicinity of $K$ near the Fermi level. The dashed line is the Fermi level; pale yellow shading indicates only the reference thermal window $|\mathcal E-\mu|\leq3k_{\mathrm B}T$.}
\label{fig:preliminary-anomalous-bands}
\end{figure*}

\subsection{Self-consistent phases and order parameters}
\label{sec:reconstructed-order}

We study the ferromagnetic, chiral, and mixed HF solutions at filling $n=1.5$, using the positive hopping $t$ as the energy unit and including neither bare spin--orbit coupling nor an external Zeeman field.

\Fig[(a)]{fig:preliminary-order} shows the $U$--$V$ phase diagram.
In the fixed spin basis, we classify the solutions using the $z$-component magnetization $m$ and the relative phases of the spin-flip bond fields $C_\ell=C_{\uparrow\downarrow,\ell}$.
Chiral bond order has at least one pair of nonzero bond fields with a relative phase different from $0$ and $\pi$.
The ferromagnetic (FM) state has $m\ne0$ without chiral bond order, whereas the chiral state has $m=0$ and nontrivial relative bond phases.
The mixed state has both finite $m$ and chiral bond order. We label the remaining weak-coupling region PM (paramagnetic metal). In the PM state, the $z$-component magnetization $m$, onsite spin coherence $\xi$, and spin-flip bond fields $C_{\uparrow\downarrow,\ell}$ are all zero, while the spin-conserving bond fields $C_{\sigma\sigma,\ell}$ are independent of spin and bond direction. Its spin-degenerate bands cross the Fermi level.

We fix $V=3t$ and scan $2<U/t<6$ to study the order parameters in the chiral, mixed, and FM states.
In the chiral regime, the $z$-component magnetization $m$ and onsite spin coherence $\xi$ in \Fig[(b)]{fig:preliminary-order} are both zero. The spin-flip bond magnitudes $|C_{\uparrow\downarrow,\ell}|$ are equal [\Fig[(d)]{fig:preliminary-order}], with successive phase differences of $2\pi/3$ modulo $2\pi$ [\Fig[(e)]{fig:preliminary-order}].

In the low-$U$ chiral state, unequal spin-conserving bond components $C_{\sigma\sigma,\ell}$ reflect different projections of the bond spin vectors onto the fixed spin axis. The spin-averaged bond fields $(C_{\uparrow\uparrow,\ell}+C_{\downarrow\downarrow,\ell})/2$ are direction-independent, while the bond spin vectors are coplanar and sum to zero. Thus, no band gap opens.

In the mixed regime, $|m|$ becomes finite and grows, while $\xi$ also develops. The spin-conserving bond fields $C_{\sigma\sigma,\ell}$ in \Fig[(c)]{fig:preliminary-order} differ between spin components and among bond directions. The spin-flip bond magnitudes $|C_{\uparrow\downarrow,\ell}|$ decrease and become unequal, while their relative phases remain nontrivial.

In the FM regime, $|m|$ varies only weakly, whereas the magnitude of $\xi$ continues to increase. The spin-conserving bond fields $C_{\sigma\sigma,\ell}$ become approximately equivalent along the three directions within each spin component; the difference between the spin-up and spin-down fields, $C_{\uparrow\uparrow,\ell}-C_{\downarrow\downarrow,\ell}$, persists and modifies the band dispersion. The spin-flip bond magnitudes $|C_{\uparrow\downarrow,\ell}|$ also become approximately equal, and their relative phases approach zero modulo $2\pi$.

The order parameters also indicate how the Bloch geometry changes along this scan (\Sec{sec:two-band-chirality}).
In the chiral state ($m=\xi=0$) and the directionally equivalent, phase-aligned FM limit, the spin directions lie in a common plane, giving zero Berry curvature away from band touchings.
The nontrivial relative bond phases can still make the phase of the spin-flip field $\Delta_{\uparrow\downarrow}(\bm k)$ depend on momentum, allowing a finite quantum metric.
In the mixed state, finite $z$-component magnetization $m$ coexists with nontrivial relative bond phases. Together, these fields reconstruct the Bloch eigenstates and allow finite Berry curvature.

\begin{figure}[!htbp]
\setlength{\abovecaptionskip}{3pt}
\centering
\includegraphics[width=\columnwidth]{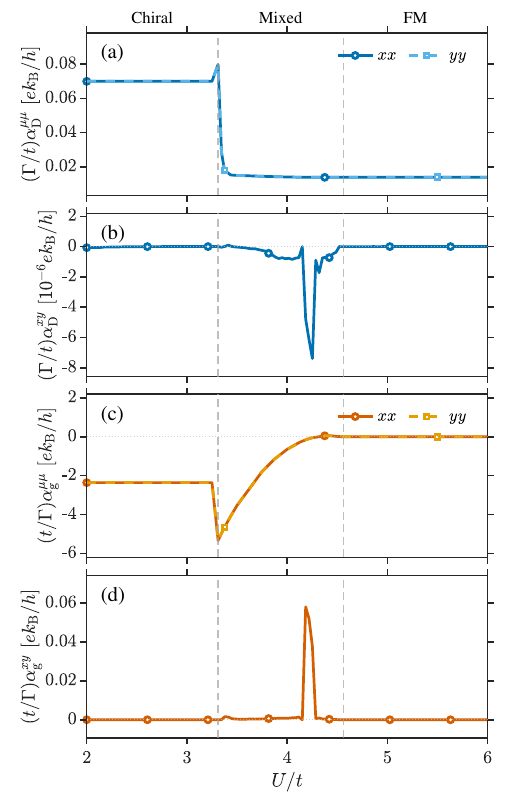}
\caption{Interaction dependence of the Drude and quantum-metric thermoelectric contributions at $n=1.5$, $k_{\mathrm B}T/t=0.01$, fixed $V/t=3$, and phenomenological broadening $\Gamma/t=0.1$.
\textbf{(a)} Longitudinal and \textbf{(b)} symmetric transverse Drude coefficients multiplied by $\Gamma/t$.
\textbf{(c)} Longitudinal and \textbf{(d)} symmetric transverse quantum-metric coefficients multiplied by $t/\Gamma$.
In the figure labels, $\mathrm D$ and $\mathrm g$ denote the Drude and quantum-metric contributions, respectively.}
\label{fig:preliminary-scaled-channels}
\end{figure}

\subsection{Anomalous thermoelectric response and band reconstruction}
\label{sec:reconstructed-geometry}
\label{sec:reconstructed-anomalous}

For $h_0(\bm k)=\epsilon_{\bm k}^{(0)}I$, we can choose a momentum-independent spin basis with $A^\mu=0$.
The bare and coherence terms then vanish, so the anomalous thermoelectric response arises entirely from the interaction-induced term.
The numerical decomposition in \Fig[(a)]{fig:preliminary-anomalous-bands} agrees with this prediction: the total response coincides with the interaction-induced contribution, while the bare and coherence contributions remain zero.
The response is concentrated in the mixed regime, where $z$-component magnetization $m$ and chiral bond order coexist, and changes from positive to negative as $U$ increases.

\Fig[(d)]{fig:preliminary-anomalous-bands} shows the direct band gap, $\Delta_{\mathrm{dir}}=\min_{\bm k}[\mathcal E_{\bm k,+}-\mathcal E_{\bm k,-}]$.
The separation opens upon entering the mixed state and closes again in the directionally equivalent, phase-aligned FM limit.
The spectrum of the chiral state in \Fig[(b)]{fig:preliminary-anomalous-bands} has a Dirac-like near-touching at $K$ above the Fermi level.
The spin directions of the Bloch eigenstates lie in a common plane, giving zero Berry curvature away from band touchings. 
In the spectrum of the FM state in \Fig[(c)]{fig:preliminary-anomalous-bands}, the upper band crosses the Fermi level, while the Berry curvature is again suppressed as the bond fields become phase-aligned and directionally equivalent.

The sign reversal follows the change in the band-resolved contributions to Eq.~\eqref{eq:entropy-weighted-alpha}. As the lower-band maximum near $K$ moves from above the Fermi level [\Fig[(e)]{fig:preliminary-anomalous-bands}] to below it [\Fig[(f)]{fig:preliminary-anomalous-bands}], the lower band becomes more fully occupied. The factor $S(f)$ tends to zero for a fully occupied state, so the positive lower-band contribution decreases as these states move out of the thermal window. The negative contribution from the partially occupied upper band then becomes dominant. Interaction-driven shifts and reconstruction of the bands alter the Berry-curvature contributions to thermoelectric transport near the Fermi level, leading to the sign reversal.

\begingroup\clubpenalty=10000
To determine whether this sign change accompanies a change in band topology, we evaluate $\mathcal C_n=(2\pi)^{-1}\int_{\mathrm{BZ}}d^2k\,\Omega_n^{xy}(\bm k)$ using the two-band Berry curvature in Eq.~\eqref{eq:appendix-two-band-projector-geometry}. For two isolated bands, $\Omega_-^{xy}=-\Omega_+^{xy}$, and hence $\mathcal C_-+\mathcal C_+=0$. Inversion acts as the identity in spin space, so $\bm k$ and $-\bm k$ map to the same point on the Bloch sphere. Since two-dimensional inversion preserves orientation, the coverings occur in pairs and each isolated band has an even Chern number. The numerical integrals for both selected mixed states give $\mathcal C_-=2$ and $\mathcal C_+=-2$, corresponding to a Bloch-sphere map with degree two in magnitude. Inversion constrains the parity of these integers, while their values follow from the reconstructed eigenstates. Their equality on the two sides of the thermoelectric sign reversal shows that the reversal does not require a change in band topology: the Chern number integrates curvature over a complete band, whereas the thermoelectric response emphasizes states in the thermal window near the Fermi level.

\par\endgroup


\begingroup
\subsection{Rotational symmetry of the Drude and quantum-metric responses}
\label{sec:reconstructed-symmetric}

\Fig{fig:preliminary-scaled-channels} compares the longitudinal $xx$ and $yy$ components and the symmetric transverse $xy$ component of the Drude contribution $\alpha_{\mathrm{intra}}$ and the quantum-metric contribution $\alpha_{\mathrm{inter,s}}$ at $\Gamma/t=0.1$.
We plot $(\Gamma/t)\alpha_{\mathrm{intra}}$ and $(t/\Gamma)\alpha_{\mathrm{inter,s}}$ to show their variation with $U$.
\par\endgroup

In the low-$U$ chiral regime, the spin-flip bond fields $C_{\uparrow\downarrow,\ell}$ and Bloch eigenstates vary little with $U$. At fixed filling, the uniform Hartree shift is absorbed into the chemical potential and does not change the band energies relative to it. The longitudinal responses remain nearly constant, with a finite quantum-metric contribution.
As discussed in \Sec{sec:two-band-chirality}, the nontrivial bond phases allow the relative phase of the two spin components to vary with momentum, producing a finite quantum metric even when the Berry curvature vanishes. In the chiral state, a compensating spin rotation preserves $C_6$ symmetry of the charge response. The longitudinal $xx$ and $yy$ responses are therefore equal, and the symmetric transverse response vanishes in both the Drude and quantum-metric channels.

In the mixed regime, finite $z$-component magnetization $m$ coexists with nontrivial relative bond phases, reconstructing both the dispersion and the Bloch states. The longitudinal Drude response briefly increases near the onset of this regime, then drops rapidly and varies more slowly at larger $U$. The magnitude of the longitudinal metric response also increases near the onset and then generally decreases toward the FM state.
As the magnetization magnitude $|m|$ grows and the spin-flip bond magnitudes $|C_{\uparrow\downarrow,\ell}|$ decrease, the onsite exchange field increasingly favors a common spin direction within each band. This tendency is consistent with the decrease in the quantum-metric response.

In the mixed state, a reduction of the charge-response rotational symmetry from $C_6$ to $C_2$ allows nonzero symmetric transverse responses in the Drude and quantum-metric channels.

At larger $U$, the system enters the FM regime, where the bond fields become directionally equivalent and their phases align. The mean fields vary more slowly, and the longitudinal Drude response approaches a constant. The strongly suppressed longitudinal metric response is consistent with the nearly aligned spin directions within each band.
The FM state restores $C_6$ symmetry of the charge response, which enforces equal longitudinal components and zero symmetric transverse response in both channels.

\subsection{Broadening dependence in the phenomenological relaxation model}
\label{sec:reconstructed-total}

To examine the broadening dependence at $U/t=4.2$ and $V/t=3$, we vary $\Gamma$ while holding the mean fields fixed, assuming that the band structure and Bloch geometry remain unchanged. The Drude and quantum-metric contributions are evaluated from Eqs.~\eqref{eq:finite-Gamma-Drude} and \eqref{eq:appendix-full-gamma-metric}, respectively.

\begin{figure}[!htbp]
\setlength{\abovecaptionskip}{3pt}
\centering
\includegraphics[width=\columnwidth]{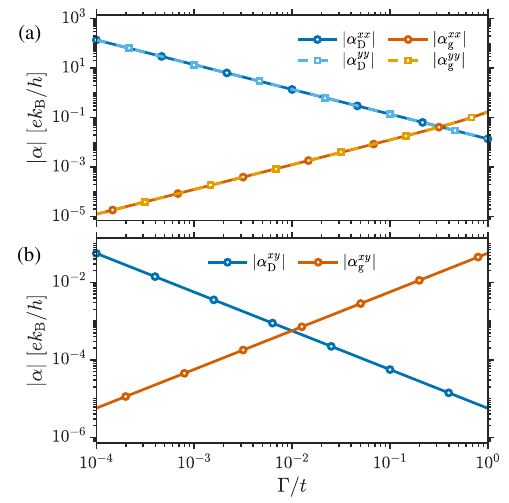}
\caption{Broadening dependence of the Drude ($\mathrm D$) and quantum-metric ($\mathrm g$) thermoelectric contributions in the mixed state.
Parameters are $U/t=4.2$, $V/t=3$, $n=1.5$, and $k_{\mathrm B}T/t=0.01$.
\textbf{(a)} Longitudinal $xx$ and $yy$ components. \textbf{(b)} Symmetric transverse $xy$ component.
The absolute values of the responses are plotted on logarithmic axes over $\Gamma/t=10^{-4}$--$1$.}

\label{fig:preliminary-actual-channels}
\end{figure}

Over the broad range of $\Gamma$ considered in \Fig{fig:preliminary-actual-channels}, the quantum-metric contribution increases roughly linearly with $\Gamma$, while the Drude contribution decreases as $\Gamma^{-1}$. At larger broadening, the quantum-metric contribution can exceed the Drude contribution.

Beyond the plotted range, once $\Gamma$ greatly exceeds the relevant interband separations, Eq.~\eqref{eq:appendix-full-gamma-metric} predicts that the quantum-metric contribution eventually decreases as $\Gamma^{-1}$ at fixed mean fields. The weak-broadening linear growth cannot persist at arbitrarily strong damping.

\begingroup
\interlinepenalty=10000
\section{Discussion}
\label{sec:discussion}

Including field-induced changes of the order parameters and their collective dynamics requires vertex corrections consistent with the self-energy \cite{BaymKadanoff1961} and with energy conservation \cite{Kontani2003}.
\par\endgroup

Triangular-lattice Hubbard and extended Hubbard models support competing orders at different fillings and interaction strengths, including $120^\circ$ antiferromagnetism, multisublattice magnetic order, and magnetic order with charge modulation \cite{MartinBatista2008,PasrijaKumar2016,PanWuDasSarma2020}.
We focus on $z$-component magnetization $m$, chiral bond order, and their coexistence.
The states in the phase diagram are obtained from restricted HF solutions that preserve translation and inversion symmetries. Including additional candidate states may change the phase diagram.
Future work can examine states with larger magnetic or charge-order unit cells and their transverse thermoelectric responses. Material-specific calculations also require realistic hopping and interaction parameters (see Appendix~\ref{app:model-checks-materials}).

A microscopic treatment of disorder should account for both its effect on the state and the associated transport processes. Scattering can produce damping that depends on momentum and band index, while disorder can change the self-consistent onsite and bond fields, and hence the spectrum, Bloch geometry, and symmetry. Determining the range over which the mixed state and its transverse responses persist therefore requires a specified disorder model incorporated into the HF self-consistency equations. The response must also include side-jump and skew-scattering contributions \cite{Nagaosa2010,Onoda2008}, which can substantially affect anomalous thermoelectric transport \cite{Weischenberg2013,PapajFu2021}; side-jump terms can share the leading $\Gamma^0$ behavior of the intrinsic Berry-curvature contribution.

\begingroup
\section{Conclusion}
\label{sec:conclusion}

We have derived the Drude, quantum-metric, and Berry-curvature contributions to linear thermoelectric transport within the HF approximation.
In the noninteracting Bloch eigenbasis, the interaction connection describes the change in the Berry connection caused by the HF self-energy. The anomalous response separates into a bare geometric term, a coherence term coupling the two connections, and a term built solely from the interaction connection.
\par\endgroup

Time-reversal symmetry forbids the antisymmetric Berry-curvature response. For the Drude and quantum-metric channels, an in-plane rotational symmetry of order greater than two in the charge response enforces longitudinal isotropy and forbids symmetric transverse transport.
In the relaxation-time approximation, the leading weak-broadening behavior is $\alpha_{\mathrm{intra}}\propto\Gamma^{-1}$, $\alpha_{\mathrm{inter,s}}\propto\Gamma$, and $\alpha_{\mathrm{inter,a}}\propto\Gamma^0$. The interband scalings require broadening smaller than the relevant band separations.
The triangular-lattice extended Hubbard model provides an example of this mechanism: onsite and bond orders generate Berry curvature and a quantum metric and can change the symmetry of the system, thereby contributing to transverse thermoelectric transport.

\begingroup
\interlinepenalty=10000
\begin{acknowledgments}
This work is supported by the National Key R\&D Program of China (Grant No.~2024YFA1409200 and No.~2022YFA1402802) and the CAS Project for Young Scientists in Basic Research (Grant No.~YSBR-057).
G.S.
was supported in part by the Quantum Science and Technology--National Science and Technology Major Project under Grant No.~2024ZD0300500, NSFC Nos.~12534009 and 12447101, the Strategic Priority Research Program of CAS (Grant No.~XDB1270000), and the CAS Superconducting Research Project under Grant No.~SCZX-0101.
\end{acknowledgments}
\endgroup

\appendix

\section{Hartree--Fock formulation and Bloch geometry}
\label{app:hf-and-bloch-geometry}

\subsection{HF self-energy and basis transformation}
\label{app:hartree-fock-details}

After Fourier transformation and a change from the localized-orbital basis to the noninteracting Bloch basis, the interaction is
\begin{equation}
\begin{aligned}
H_{\mathrm{int}}=&\frac{1}{2N_{\mathrm c}}
\sum_{\{\bm k_i,a_i\}_{i=1}^{4}}
V_{a_1a_2a_3a_4}(\bm k_1,\bm k_2,\bm k_3,\bm k_4)\\
&\times c_{\bm k_1a_1}^{\dagger}c_{\bm k_2a_2}^{\dagger}
c_{\bm k_3a_3}c_{\bm k_4a_4}.
\end{aligned}
\label{eq:appendix-general-Hint}
\end{equation}
The vertex enforces crystal-momentum conservation, $\bm k_1+\bm k_2=\bm k_3+\bm k_4+\bm G$.
For a symmetric two-particle interaction it can be chosen to obey
\begin{equation}
\begin{aligned}
&V_{a_1a_2a_3a_4}(\bm k_1,\bm k_2,\bm k_3,\bm k_4)\\
&\quad=V_{a_2a_1a_4a_3}(\bm k_2,\bm k_1,\bm k_4,\bm k_3).
\end{aligned}
\label{eq:appendix-vertex-pair-exchange}
\end{equation}

For compactness in the Wick decomposition, let $1$ denote $(\bm k_1,a_1)$, and similarly for the remaining labels.
Retaining the one-body and constant terms in the HF decoupling gives
\begin{equation}
\begin{aligned}
c_1^\dagger c_2^\dagger c_3c_4\simeq{}
&\langle c_1^\dagger c_4\rangle c_2^\dagger c_3
+\langle c_2^\dagger c_3\rangle c_1^\dagger c_4\\
&-\langle c_1^\dagger c_3\rangle c_2^\dagger c_4
-\langle c_2^\dagger c_4\rangle c_1^\dagger c_3\\
&-\langle c_1^\dagger c_4\rangle\langle c_2^\dagger c_3\rangle
+\langle c_1^\dagger c_3\rangle\langle c_2^\dagger c_4\rangle.
\end{aligned}
\label{eq:appendix-full-wick-decoupling}
\end{equation}
Translation symmetry makes the reduced density matrix diagonal in momentum,
\begin{equation}
\langle c_{\bm k'b}^{\dagger}c_{\bm k a}\rangle
=\delta_{\bm k\bm k'}\rho_{ab}(\bm k).
\label{eq:appendix-momentum-diagonal-density-matrix}
\end{equation}
The density matrix itself is defined as
\begin{equation}
\rho_{ab}(\bm k)=
\langle c_{\bm k b}^{\dagger}c_{\bm k a}\rangle.
\label{eq:appendix-reference-density-matrix}
\end{equation}

Using the pair-exchange property, the two equal direct contractions cancel the factor $1/2$ in Eq.~\eqref{eq:appendix-general-Hint}.
The resulting Hartree self-energy is
\begin{equation}
\Sigma^{\mathrm H}_{ab}(\bm k)=
\frac{1}{N_{\mathrm c}}\sum_{\bm k'}\sum_{c,d}
V_{adcb}(\bm k,\bm k',\bm k',\bm k)\rho_{cd}(\bm k').
\label{eq:appendix-hartree-self-energy}
\end{equation}
The two equal exchange contractions behave in the same way, while fermionic exchange supplies a minus sign.
Hence
\begin{equation}
\Sigma^{\mathrm F}_{ab}(\bm k)=
-\frac{1}{N_{\mathrm c}}\sum_{\bm k'}\sum_{c,d}
V_{adbc}(\bm k,\bm k',\bm k,\bm k')\rho_{cd}(\bm k').
\label{eq:appendix-fock-self-energy}
\end{equation}
The products of expectation values yield the double-counting constant
\begin{equation}
E_{\mathrm{const}}[\rho]=-
\frac{1}{2}\sum_{\bm k}\sum_{a,b}
[\Sigma^{\mathrm H}_{ab}(\bm k)+\Sigma^{\mathrm F}_{ab}(\bm k)]
\rho_{ba}(\bm k).
\label{eq:appendix-mean-field-constant}
\end{equation}
Together, Eqs.~\eqref{eq:appendix-hartree-self-energy}--\eqref{eq:appendix-mean-field-constant} produce the Hamiltonian in Eqs.~\eqref{eq:general-HMF} and \eqref{eq:mean-field-single-particle-matrix}.

The connection in the noninteracting basis is $A_{ab}^\mu=i\langle a(\bm k)|\partial_{k_\mu}|b(\bm k)\rangle$, while the connection in the interacting basis is $\mathcal A_{mn}^\mu=i\langle m(\bm k)|\partial_{k_\mu}|n(\bm k)\rangle$.
The transformation $c_{\bm k a}=\sum_nW_{an}d_{\bm k n}$ gives
\begin{equation}
\begin{aligned}
\mathcal A_{nm}^{\mu}={}&
\sum_{a,b}W_{an}^{*}A_{ab}^{\mu}W_{bm}
+i\sum_aW_{an}^{*}\partial_{k_\mu}W_{am}.
\end{aligned}
\label{eq:appendix-connection-transformation}
\end{equation}

\subsection{Two-band Berry curvature and quantum metric}
\label{app:mean-field-quantum-metric}

For $h=d_0I+\bm d\cdot\bm\sigma$, define $D=|\bm d|$ and $\hat{\bm d}=\bm d/D$. The projectors $P_\pm=(I\pm\hat{\bm d}\cdot\bm\sigma)/2$ give
\begin{equation}
 \begin{aligned}
 g_\pm^{\mu\nu}
 &=\frac14\,\partial_{k_\mu}\hat{\bm d}
               \cdot\partial_{k_\nu}\hat{\bm d},\\
 \Omega_\pm^{\mu\nu}
 &=\mp\frac12\,\hat{\bm d}\cdot
 (\partial_{k_\mu}\hat{\bm d}
       \times\partial_{k_\nu}\hat{\bm d}).
 \end{aligned}
 \label{eq:appendix-two-band-projector-geometry}
\end{equation}
These follow by evaluating $\operatorname{Tr}[P_\pm(\partial_{k_\mu}P_\pm)(\partial_{k_\nu}P_\pm)]$. For Eq.~\eqref{eq:generic-constant-diagonal-model}, $\bm d=(\operatorname{Re}\mathcal D,\operatorname{Im}\mathcal D,m_0)$ and $D^2=m_0^2+|\mathcal D|^2$. Substitution gives Eq.~\eqref{eq:generic-constant-mass-curvature} and
\begin{equation}
 \begin{aligned}
 g_{+-}^{\mu\nu}=\frac{1}{4D^4}\bigl\{&
 m_0^2\operatorname{Re}
 [\partial_{k_\mu}\mathcal D^*\partial_{k_\nu}\mathcal D]\\
 &+\operatorname{Im}[\mathcal D^*\partial_{k_\mu}\mathcal D]\,
   \operatorname{Im}[\mathcal D^*\partial_{k_\nu}\mathcal D]\bigr\}.
 \end{aligned}
 \label{eq:appendix-mf-metric-complex}
\end{equation}
Here $g_+^{\mu\nu}=g_-^{\mu\nu}=g_{+-}^{\mu\nu}=g_{-+}^{\mu\nu}$. For the cosine form in Eq.~\eqref{eq:generic-cosine-order}, differentiation gives
\begin{equation}
 \partial_{k_\mu}\mathcal D
 =-2\Lambda\sum_\ell\mathcal C_\ell\eta_{\ell\mu}
            \sin(\bm k\cdot\bm\eta_\ell).
 \label{eq:appendix-complex-field-gradients}
\end{equation}
Combining the $(\ell,\ell')$ and $(\ell',\ell)$ terms in the gradient product yields Eq.~\eqref{eq:generic-pairwise-chirality}.

Writing $\mathcal D=|\mathcal D|e^{i\varphi}$ gives $\operatorname{Im}(\mathcal D^*\partial_{k_\mu}\mathcal D)=|\mathcal D|^2\partial_{k_\mu}\varphi$. In the planar limit $m_0=0$, at momenta where $|\mathcal D|\ne0$, the metric therefore reduces to
\begin{equation}
 g_\pm^{\mu\nu}=\frac14
 (\partial_{k_\mu}\varphi)(\partial_{k_\nu}\varphi).
 \label{eq:appendix-planar-metric}
\end{equation}
The symmetry constraints on the resulting transport integrals are discussed in Secs.~\ref{sec:scattering-symmetry} and \ref{sec:two-band-chirality}.

If $\hat{\bm d}(\bm k)$ lies in a fixed plane through the origin, the scalar triple product vanishes, while its derivatives can still give a finite metric. This explains the coplanar limit without invoking time-reversal symmetry and applies away from band touchings. More generally,
\begin{equation}
 g_\pm^{xx}g_\pm^{yy}-(g_\pm^{xy})^2
 =\frac{(\Omega_\pm^{xy})^2}{4},
 \label{eq:appendix-two-band-metric-curvature-identity}
\end{equation}
because both derivatives are tangent to the unit sphere. Nonzero curvature therefore requires a nonzero metric.

\section{Density-matrix solution and Kubo response}
\label{app:thermoelectric-response}

We first obtain the density-matrix response and the resulting local Kubo current. The conversion to transport coefficients, including orbital magnetization and finite broadening, is given in Appendix~\ref{app:transport-coefficients}.

\subsection{Thermal-field density-matrix solution}
\label{app:density-matrix-details}

The Hermitian thermal perturbation in the one-particle Hilbert space is
\begin{equation}
h_T(t)=-\frac{1}{2}\{\widetilde h_{\mathrm{MF}},
\bm r\cdot\bm E_T(t)\}.
\label{eq:appendix-single-particle-thermal-perturbation}
\end{equation}
In second quantization it becomes
\begin{equation}
H_T(t)=-\frac{1}{2}E_T^\mu(t)
\sum_{\bm k,m,n}[q_{\bm k}^{\mu}]_{mn}
d_{\bm k m}^{\dagger}d_{\bm k n},
\label{eq:appendix-second-quantized-thermal-perturbation}
\end{equation}
where $q_{\bm k}^{\mu}=\{\widetilde h_{\mathrm{MF}},r_{\bm k}^{\mu}\}$.
Because the derivative in the position operator acts on every quantity to its right, the matrix differential operator is
\begin{equation}
\begin{aligned}
\relax[q_{\bm k}^{\mu}]_{mn}={}&
2i\widetilde{\mathcal E}_{\bm k m}\delta_{mn}\partial_{k_\mu}
+i(\partial_{k_\mu}\widetilde{\mathcal E}_{\bm k m})\delta_{mn}\\
&+(\widetilde{\mathcal E}_{\bm k m}
+\widetilde{\mathcal E}_{\bm k n})\mathcal A_{mn}^{\mu}.
\end{aligned}
\label{eq:appendix-thermal-anticommutator}
\end{equation}

For $H(t)=H_{\mathrm{MF}}+H_T(t)$, the reduced density matrix obeys
\begin{equation}
i\hbar\partial_t\rho_{mn}
=\mathcal E_{\bm k,mn}\rho_{mn}
-\frac{1}{2}E_T^\mu(t)[q_{\bm k}^{\mu},\rho]_{mn}.
\label{eq:appendix-reduced-liouville}
\end{equation}
Using $\rho(t)=\int d\omega\,e^{-i\omega t}\rho(\omega)/(2\pi)$ and expanding in powers of the thermal field gives
\begin{equation}
 \begin{aligned}
 &(\hbar\omega-\mathcal E_{\bm k,mn}+i\eta)
       \rho_{mn}^{(p+1)}(\bm k,\omega)\\
 &\quad=-\frac12\int\frac{d\omega_1}{2\pi}E_T^\mu(\omega_1)
 [q_{\bm k}^{\mu},\rho^{(p)}(\bm k,\omega-\omega_1)]_{mn}.
 \end{aligned}
 \label{eq:appendix-frequency-liouville}
\end{equation}
The retarded boundary condition supplies $\eta=0^+$. Dividing by the factor on the left gives Eq.~\eqref{eq:thermal-density-matrix-recursion-main}.

In evaluating $[q_{\bm k}^{\mu},f]$, the derivative part of $q_{\bm k}^{\mu}$ acts on both $f$ and the band spinor.
Acting on an arbitrary band spinor $\psi$ gives
\begin{equation}
\begin{aligned}
([q_{\bm k}^{\mu},f]\psi)_m={}&
2i\widetilde{\mathcal E}_{\bm k m}
(\partial_{k_\mu}f_{\bm k m})\psi_m\\
&+\sum_n(\widetilde{\mathcal E}_{\bm k m}
+\widetilde{\mathcal E}_{\bm k n})\mathcal A_{mn}^{\mu}
(f_{\bm k n}-f_{\bm k m})\psi_n.
\end{aligned}
\label{eq:appendix-thermal-commutator-on-spinor}
\end{equation}
The terms proportional to $\partial_{k_\mu}\psi_m$ and $\partial_{k_\mu}\widetilde{\mathcal E}_{\bm k m}$ cancel pairwise.
Thus the required matrix elements are
\begin{equation}
\begin{aligned}
\relax[q_{\bm k}^{\mu},f]_{mn}={}&
2i\widetilde{\mathcal E}_{\bm k m}
(\partial_{k_\mu}f_{\bm k m})\delta_{mn}\\
&+(\widetilde{\mathcal E}_{\bm k m}
+\widetilde{\mathcal E}_{\bm k n})\mathcal A_{mn}^{\mu}
(f_{\bm k n}-f_{\bm k m}).
\end{aligned}
\label{eq:appendix-thermal-driving-commutator}
\end{equation}
Substituting Eq.~\eqref{eq:appendix-thermal-driving-commutator} into the recursion relation separately for $m=n$ and $m\ne n$ gives Eqs.~\eqref{eq:rho-diagonal} and \eqref{eq:rho-offdiagonal}.

\subsection{Intraband contribution}

The commutator $v^\mu=[r^\mu,h_{\mathrm{MF}}]/(i\hbar)$, with $r^\mu=i\partial_{k_\mu}+\mathcal A^\mu$, gives
\begin{equation}
 v_{\bm k,mn}^{\mu}
 =\frac{\partial_{k_\mu}\mathcal E_{\bm k m}}{\hbar}\delta_{mn}
 +\frac{i}{\hbar}\mathcal E_{\bm k,mn}\mathcal A_{mn}^{\mu}.
 \label{eq:appendix-velocity-matrix}
\end{equation}
The derivative acts on the band energy, whereas the connection commutator supplies the off-diagonal energy difference.

The diagonal density-matrix response produces the intraband Kubo coefficient before integration by parts,
\begin{equation}
L_{\mathrm{intra}}^{\mathrm{Kubo},\mu\nu}(\omega)=
\frac{ie}{\hbar\omega+i\eta}\intk\sum_n
\widetilde{\mathcal E}_{\bm k n}v_{\bm k n}^{\mu}
\partial_{k_\nu}f_{\bm k n}.
\label{eq:appendix-intraband-before-parts}
\end{equation}
The band-diagonal heat-current matrix element is $\widetilde{\mathcal E}_{\bm k n}v_{\bm k n}^{\mu}$.
For comparison with the expanded form, define $(\widetilde m^{-1})_n^{\mu\nu}=\hbar^{-2}\partial_{k_\mu}\partial_{k_\nu}\mathcal E_n$.
Periodicity over the Brillouin zone removes the boundary term and gives the following identity, with momentum labels suppressed:
\begin{equation}
\begin{aligned}
\intk\widetilde{\mathcal E}_{n}v_n^\mu\partial_{k_\nu}f_n
={}&-\intk f_n\partial_{k_\nu}(\widetilde{\mathcal E}_n v_n^\mu)\\
={}&-\hbar\intk f_n v_n^\mu v_n^\nu\\
&-\hbar\intk f_n\widetilde{\mathcal E}_{n}
(\widetilde m^{-1})_n^{\mu\nu}.
\end{aligned}
\label{eq:appendix-intraband-integration-by-parts}
\end{equation}
Taking $\alpha_{\mathrm{intra}}^{\mu\nu}(\omega)=\operatorname{Re}L_{\mathrm{intra}}^{\mathrm{Kubo},\mu\nu}(\omega)/T$ yields Eq.~\eqref{eq:intraband-dynamical-alpha}.

\subsection{Interband decomposition}

The off-diagonal response gives
\begin{equation}
\begin{aligned}
L_{\mathrm{inter}}^{\mathrm{Kubo},\mu\nu}(\omega)
=-&\frac{i e}{2\hbar}\intk\sum_{m\ne n}
\frac{\mathcal E_{\bm k,nm}}
{\hbar\omega-\mathcal E_{\bm k,mn}+i\eta}\\
&\times(\widetilde{\mathcal E}_{\bm k m}
+\widetilde{\mathcal E}_{\bm k n})
\mathcal A_{nm}^{\mu}\mathcal A_{mn}^{\nu}f_{\bm k,mn}.
\end{aligned}
\label{eq:appendix-interband-kubo}
\end{equation}
The symmetric geometric factor defines the quantum-metric matrix element,
\begin{equation}
g_{mn}^{\mu\nu}=\frac{1}{2}
(\mathcal A_{mn}^{\nu}\mathcal A_{nm}^{\mu}
+\mathcal A_{mn}^{\mu}\mathcal A_{nm}^{\nu}).
\label{eq:appendix-pair-quantum-metric}
\end{equation}
The antisymmetric factor defines the Berry-curvature matrix element,
\begin{equation}
\Omega_{mn}^{\mu\nu}=i
(\mathcal A_{mn}^{\mu}\mathcal A_{nm}^{\nu}
-\mathcal A_{mn}^{\nu}\mathcal A_{nm}^{\mu}).
\label{eq:appendix-pair-berry-curvature}
\end{equation}
Accordingly, $\mathcal A_{nm}^{\mu}\mathcal A_{mn}^{\nu} =g_{mn}^{\mu\nu}+i\Omega_{mn}^{\mu\nu}/2$.

\subsection{Quantum-metric contribution}

For compactness, define the frequency-dependent response factor
\begin{equation}
\mathcal R_{\bm k,mn}(\omega)=
\frac{\mathcal E_{\bm k,nm}}
{\hbar\omega-\mathcal E_{\bm k,mn}+i\eta}.
\label{eq:appendix-response-factor}
\end{equation}
The symmetric part of Eq.~\eqref{eq:appendix-interband-kubo} is
\begin{equation}
\begin{aligned}
L_{\mathrm{inter,s}}^{\mathrm{Kubo},\mu\nu}(\omega)
=-&\frac{ie}{2\hbar}\intk\sum_{m\ne n}
\mathcal R_{\bm k,mn}(\omega)\\
&\times(\widetilde{\mathcal E}_{\bm k m}
+\widetilde{\mathcal E}_{\bm k n})
g_{mn}^{\mu\nu}f_{\bm k,mn}.
\end{aligned}
\label{eq:appendix-symmetric-interband-response}
\end{equation}
Its absorptive part follows from
\begin{equation}
\lim_{\eta\to0^+}\operatorname{Im}\mathcal R_{\bm k,mn}=
-\pi\mathcal E_{\bm k,nm}
\delta(\hbar\omega-\mathcal E_{\bm k,mn}).
\label{eq:appendix-clean-response-factor-imaginary}
\end{equation}
Since the real part of $-i\mathcal R$ is $\operatorname{Im}\mathcal R$, Eq.~\eqref{eq:appendix-symmetric-interband-response} first gives a single resonant delta function.
Exchanging $m$ and $n$ in the term proportional to $f_{\bm k n}$ then yields
\begin{equation}
\begin{aligned}
\operatorname{Re}L_{\mathrm{inter,s}}^{\mathrm{Kubo},\mu\nu}(\omega)
=-&\frac{\pi e}{2\hbar}\intk\sum_{m\ne n}
\mathcal E_{\bm k,nm}\\
&\times\left[\delta(\hbar\omega-\mathcal E_{\bm k,mn})
+\delta(\hbar\omega-\mathcal E_{\bm k,nm})\right]\\
&\times(\widetilde{\mathcal E}_{\bm k m}
+\widetilde{\mathcal E}_{\bm k n})
g_{mn}^{\mu\nu}f_{\bm k m}.
\end{aligned}
\label{eq:appendix-clean-metric-response}
\end{equation}
Orbital magnetization does not contribute to the symmetric interband coefficient.
Therefore $\alpha_{\mathrm{inter,s}}^{\mu\nu}=\operatorname{Re} L_{\mathrm{inter,s}}^{\mathrm{Kubo},\mu\nu}/T$, which gives Eq.~\eqref{eq:interband-symmetric-alpha-main}.
For nondegenerate bands, the delta functions have no support at zero frequency, so this clean-limit contribution vanishes.

\section{Transport coefficients and finite broadening}
\label{app:transport-coefficients}

\subsection{Transport current and orbital magnetization}
\label{app:transport-current}

The transverse local current contains equilibrium circulation that does not transport charge through the leads.
The subscripts $N$, $E$, and $Q$ denote particle, energy, and heat currents, respectively, with $\bm j_e=-e\bm j_N$ and $\bm j_Q=\bm j_E-\mu\bm j_N$ for uniform chemical potential.
The corresponding heat magnetization is $\bm M_Q=\bm M_E-\mu\bm M_N$.
In equilibrium, the particle and energy currents can be represented by their magnetization densities as
\begin{equation}
\bm j_{N(E)}^{(0)}=\bm\nabla\times\bm M_{N(E)}.
\label{eq:appendix-equilibrium-magnetization-current}
\end{equation}
To linear order in the external field, a local current is the sum of an instantaneous local-equilibrium part (le) and a Kubo response part,
\begin{equation}
\bm j_{N(Q)}=\bm j_{N(Q)}^{\mathrm{le}}
+\bm j_{N(Q)}^{\mathrm{Kubo}}.
\label{eq:appendix-local-current-decomposition}
\end{equation}
The measurable transport current is obtained by subtracting the complete field-dependent magnetization circulation,
\begin{equation}
\bm j_{N(Q)}^{\mathrm{tr}}=\bm j_{N(Q)}-
\bm\nabla\times[\bm M_{N(Q)}+\delta\bm M_{N(Q)}].
\label{eq:appendix-transport-current-subtraction}
\end{equation}

After accounting for particle magnetization, the force--flux relation for the transport particle current driven by a temperature gradient is \cite{Cooper1997,QinNiuShi2011}
\begin{equation}
 j_N^{\mathrm{tr},\mu}
 =\left[\operatorname{Re}L_{12}^{NQ,\mathrm{Kubo},\mu\nu}
       +\beta^{-1}M_N^{\mu\nu}\right]X_Q^\nu,
 \label{eq:appendix-particle-thermal-response}
\end{equation}
where $X_Q^\nu=\partial_\nu\beta=\beta E_T^\nu$ for a homogeneous reference state. Multiplying by the electron charge $-e$ and using
\begin{equation}
 \begin{aligned}
 L_{\mathrm{dc}}^{\mathrm{Kubo},\mu\nu}
 &=-e\beta L_{12}^{NQ,\mathrm{Kubo},\mu\nu},\\
 M_{\mathrm{orb}}^{\mu\nu}&=-eM_N^{\mu\nu}
 \end{aligned}
 \label{eq:appendix-current-convention-conversion}
\end{equation}
gives the electric-current convention used in the main text.

For a homogeneous sample driven only by $\bm E_T=-\bm\nabla T/T$, the electric transport current is therefore
\begin{equation}
j_e^{\mathrm{tr},\mu}=
\left[\operatorname{Re}L_{\mathrm{dc}}^{\mathrm{Kubo},\mu\nu}
+M_{\mathrm{orb}}^{\mu\nu}\right]E_T^\nu.
\label{eq:appendix-electric-thermal-field-response}
\end{equation}
Since the experimental convention is $j_e^{\mathrm{tr},\mu}=\alpha^{\mu\nu}(-\partial_\nu T)$, Eq.~\eqref{eq:appendix-electric-thermal-field-response} gives
\begin{equation}
\alpha^{\mu\nu}=\frac{1}{T}
\left[\operatorname{Re}L_{\mathrm{dc}}^{\mathrm{Kubo},\mu\nu}
+M_{\mathrm{orb}}^{\mu\nu}\right].
\label{eq:appendix-alpha-transport-definition}
\end{equation}
This is the origin of Eq.~\eqref{eq:alpha-from-kubo-and-magnetization-main}.
Orbital magnetization enters only the antisymmetric interband transport coefficient.

\subsection{Berry-curvature contribution to thermoelectric transport}
\label{app:berry-transport}

For the antisymmetric response, the static limit is taken first and the clean limit second.
Then $\operatorname{Re}\mathcal R_{\bm k,mn}(0)\to1$ and only the Berry-curvature part survives.
Exchanging the band indices $m$ and $n$ gives
\begin{equation}
\begin{aligned}
L_{\mathrm{inter,a,dc}}^{\mathrm{Kubo},\mu\nu}
={}&\frac{e}{4\hbar}\intk\sum_{m\ne n}
(\widetilde{\mathcal E}_{m}+\widetilde{\mathcal E}_{n})
\Omega_{mn}^{\mu\nu}f_{mn}\\
={}&\frac{e}{2\hbar}\intk\sum_{m\ne n}
(2\widetilde{\mathcal E}_{m}+\mathcal E_{nm})
\Omega_{mn}^{\mu\nu}f_m\\
={}&\frac{e}{\hbar}\intk\sum_{m\ne n}
\widetilde{\mathcal E}_{m}\Omega_{mn}^{\mu\nu}f_m\\
&-\intk\sum_{m\ne n}m_{\mathrm{orb},mn}^{\mu\nu}f_m.
\end{aligned}
\label{eq:appendix-antisymmetric-kubo}
\end{equation}
Momentum labels are suppressed in the last equation.
The orbital-moment matrix element used there is
\begin{equation}
m_{\mathrm{orb},mn}^{\mu\nu}=
\frac{e}{2\hbar}(\mathcal E_{\bm k m}-\mathcal E_{\bm k n})
\Omega_{mn}^{\mu\nu}.
\label{eq:appendix-orbital-moment}
\end{equation}

For the self-consistent mean-field bands, the equilibrium orbital magnetization is
\begin{equation}
\begin{aligned}
M_{\mathrm{orb}}^{\mu\nu}={}&
\intk\sum_{m\ne n}m_{\mathrm{orb},mn}^{\mu\nu}f_{\bm k m}\\
&+\frac{e}{\beta\hbar}\intk\sum_{m\ne n}
\Omega_{mn}^{\mu\nu}
\ln(1+e^{-\beta\widetilde{\mathcal E}_{\bm k m}}).
\end{aligned}
\label{eq:appendix-orbital-magnetization}
\end{equation}
Its first term cancels the orbital-moment current in Eq.~\eqref{eq:appendix-antisymmetric-kubo}.
The remaining energy and logarithmic weights combine through
\begin{equation}
\beta\widetilde{\mathcal E}f
+\ln(1+e^{-\beta\widetilde{\mathcal E}})=S(f).
\label{eq:appendix-entropy-identity}
\end{equation}
Consequently, the transport definition in Eq.~\eqref{eq:appendix-alpha-transport-definition} reduces exactly to the Berry-curvature thermoelectric response in Eq.~\eqref{eq:entropy-weighted-alpha}.
This cancellation is why the local Kubo response alone is not the anomalous thermoelectric transport coefficient.

\subsection{Finite broadening}
\label{app:finite-broadening}

Within the relaxation-time approximation used in the main text, the reduced density matrix obeys
\begin{equation}
i\hbar\partial_t\rho=
[h_{\mathrm{MF}}+h_T(t),\rho]-i\Gamma(\rho-\rho^{(0)}).
\label{eq:appendix-relaxation-liouville}
\end{equation}
The phenomenological broadening is related to the relaxation time by
\begin{equation}
\Gamma=\frac{\hbar}{\tau}.
\label{eq:appendix-gamma-tau}
\end{equation}
Replacing the retarded infinitesimal by $\Gamma$ in the dc interband response factor gives
\begin{equation}
\operatorname{Im}\mathcal R_{\bm k,mn}(0)=
-\frac{\Gamma\mathcal E_{\bm k,nm}}
{\mathcal E_{\bm k,mn}^2+\Gamma^2}.
\label{eq:appendix-finite-gamma-response-factor}
\end{equation}

Before expanding in $\Gamma$, substitution of Eq.~\eqref{eq:appendix-finite-gamma-response-factor} into Eq.~\eqref{eq:appendix-symmetric-interband-response} gives
\begin{equation}
 \begin{aligned}
 \alpha_{\mathrm{inter,s}}^{\mu\nu}(\Gamma)
 ={}&\frac{e\Gamma}{2\hbar T}\intk\sum_{m\ne n}
 \frac{\mathcal E_{\bm k,mn}}
      {\mathcal E_{\bm k,mn}^2+\Gamma^2}\\
 &\times(\widetilde{\mathcal E}_{\bm k m}
        +\widetilde{\mathcal E}_{\bm k n})
 g_{mn}^{\mu\nu}f_{\bm k,mn}.
 \end{aligned}
 \label{eq:appendix-full-gamma-metric}
\end{equation}

Away from degeneracies and for $\Gamma\ll|\mathcal E_{\bm k,mn}|$, this becomes
\begin{equation}
\operatorname{Im}\mathcal R_{\bm k,mn}(0)
\simeq\frac{\Gamma}{\mathcal E_{\bm k,mn}}.
\label{eq:appendix-weak-gamma-response-factor}
\end{equation}
Substitution into the symmetric part of Eq.~\eqref{eq:appendix-interband-kubo}, followed by exchange of $m$ and $n$ in one term, yields the local response
\begin{equation}
\operatorname{Re}L_{\mathrm{inter,s}}^{\mathrm{Kubo},\mu\nu}=
-\frac{e\Gamma}{\hbar}\intk\sum_{m\ne n}
G_{mn}^{\mu\nu}f_{\bm k m}.
\label{eq:appendix-finite-gamma-local-metric}
\end{equation}
Using the definition of $G_{mn}^{\mu\nu}$ in Eq.~\eqref{eq:thermal-connection-polarizability} and dividing by $T$ gives Eq.~\eqref{eq:finite-Gamma-metric}. Within the same relaxation-time approximation, for $\Gamma$ much larger than the relevant interband separations, the quantum-metric contribution decays as $\Gamma^{-1}$. Near a band touching, the weak-broadening expansion is not uniform; exactly degenerate subspaces require a basis-independent treatment.
The antisymmetric Berry-curvature response involves $\operatorname{Re}\mathcal R_{\bm k,mn}(0)=\mathcal E_{\bm k,mn}^2/(\mathcal E_{\bm k,mn}^2+\Gamma^2)$.
For $\Gamma\ll|\mathcal E_{\bm k,mn}|$, this factor approaches unity; retaining the leading clean-limit term and including orbital magnetization gives the $\Gamma^0$ intrinsic response used in the main text.

\section{Interaction-resolved anomalous response}
\label{app:noninteracting-basis-representation}

Momentum arguments are suppressed below when unambiguous.

For the antisymmetric interband term, $\rho=WfW^\dagger$ is the equilibrium density matrix in the noninteracting basis and $f=\operatorname{diag}(f_{\bm k n})$.
The entropy matrix is defined by
\begin{equation}
S(\rho)=-\rho\ln\rho-(I-\rho)\ln(I-\rho)=WS(f)W^\dagger.
\label{eq:appendix-entropy-matrix}
\end{equation}
The element $[\ln\rho]_{ab}$ is obtained by evaluating the matrix function $\ln\rho$ and taking its $ab$ entry.
With $\widetilde{\mathcal A}^{\mu}=W\mathcal A^{\mu}W^\dagger$, the band sum in Eq.~\eqref{eq:entropy-weighted-alpha} becomes
\begin{equation}
\begin{aligned}
\sum_{m\ne n}\Omega_{mn}^{\mu\nu}S(f_m)
={}&i\sum_{a,b}S(\rho)_{ba}
\sum_c(\widetilde{\mathcal A}_{ac}^{\mu}
\widetilde{\mathcal A}_{cb}^{\nu}
-\widetilde{\mathcal A}_{ac}^{\nu}
\widetilde{\mathcal A}_{cb}^{\mu})\\
={}&i\operatorname{Tr}
[S(\rho)[\widetilde{\mathcal A}^{\mu},
\widetilde{\mathcal A}^{\nu}]].
\end{aligned}
\label{eq:appendix-curvature-trace-transformation}
\end{equation}
The connection transformation in Eq.~\eqref{eq:appendix-connection-transformation} is equivalently $\widetilde{\mathcal A}^{\mu}=A^\mu+\mathcal I^\mu$, with $\mathcal I^\mu=i(\partial_{k_\mu}W)W^\dagger$.
Its commutator expands as
\begin{equation}
\begin{aligned}
\relax[\widetilde{\mathcal A}^{\mu},\widetilde{\mathcal A}^{\nu}]
={}&[A^\mu,A^\nu]+[A^\mu,\mathcal I^\nu]\\
&+[\mathcal I^\mu,A^\nu]+[\mathcal I^\mu,\mathcal I^\nu].
\end{aligned}
\label{eq:appendix-connection-commutator-expansion}
\end{equation}
Equations~\eqref{eq:appendix-curvature-trace-transformation} and \eqref{eq:appendix-connection-commutator-expansion} prove the three-term decomposition in Eqs.~\eqref{eq:interaction-resolved-alpha}--\eqref{eq:interaction-curvature-contribution}.

\section{Symmetry constraints on thermoelectric transport}
\label{app:symmetry-details}

\subsection{General symmetry constraints}

An in-plane proper rotation is represented by
\begin{equation}
R(\vartheta)=
\begin{pmatrix}
\cos\vartheta&-\sin\vartheta\\
\sin\vartheta&\cos\vartheta
\end{pmatrix}.
\label{eq:appendix-rotation-matrix}
\end{equation}
If an in-plane rotation is a symmetry of the state, invariance of the constitutive relation for the polar vectors $\bm j_e^{\mathrm{tr}}$ and $-\bm\nabla T$ requires
\begin{equation}
\alpha=R(\vartheta)\alpha R^T(\vartheta).
\label{eq:appendix-rotation-invariance}
\end{equation}
For the symmetric tensor, this is equivalent to
\begin{equation}
\begin{pmatrix}
\frac{1}{2}\sin2\vartheta&\cos2\vartheta-1\\[2pt]
\frac{1-\cos2\vartheta}{2}&\sin2\vartheta
\end{pmatrix}
\begin{pmatrix}
\alpha^{xx}-\alpha^{yy}\\
\alpha_{\mathrm{s}}^{xy}
\end{pmatrix}=0.
\label{eq:appendix-symmetric-rotation-constraint}
\end{equation}
The determinant is $1-\cos2\vartheta$.
Therefore any proper rotation with order $n>2$ enforces $\alpha^{xx}=\alpha^{yy}$ and $\alpha_{\mathrm{s}}^{xy}=0$, whereas $C_{2z}$ imposes no such condition.
The antisymmetric two-dimensional tensor is invariant under every proper rotation.
Reflection in the $yz$ plane acts on the in-plane coordinates through
\begin{equation}
\mathcal M_x=\begin{pmatrix}-1&0\\0&1\end{pmatrix}.
\label{eq:appendix-mirror-matrix}
\end{equation}
Mirror invariance requires
\begin{equation}
\alpha=\mathcal M_x\alpha \mathcal M_x^T.
\label{eq:appendix-mirror-invariance}
\end{equation}
Thus, $\alpha^{xy}=\alpha^{yx}=0$ in the specified $x,y$ coordinates.

The inversion and time-reversal partners of band $m$ are denoted by $Pm$ and $\bar m$, respectively.
For geometric matrix elements, inversion gives the Berry-curvature transformation
\begin{equation}
\Omega_{Pm,Pn}^{\mu\nu}(-\bm k)
=+\Omega_{mn}^{\mu\nu}(\bm k).
\label{eq:appendix-berry-inversion}
\end{equation}
Time reversal gives
\begin{equation}
\Omega_{\bar m,\bar n}^{\mu\nu}(-\bm k)
=-\Omega_{mn}^{\mu\nu}(\bm k).
\label{eq:appendix-berry-time-reversal}
\end{equation}
The quantum metric is even under inversion,
\begin{equation}
g_{Pm,Pn}^{\mu\nu}(-\bm k)
=+g_{mn}^{\mu\nu}(\bm k).
\label{eq:appendix-metric-inversion}
\end{equation}
It is also even under time reversal,
\begin{equation}
g_{\bar m,\bar n}^{\mu\nu}(-\bm k)
=+g_{mn}^{\mu\nu}(\bm k).
\label{eq:appendix-metric-time-reversal}
\end{equation}
The quasiparticle energies, Fermi functions, and factors $S(f)$ are even under these mappings.

For a combined antiunitary symmetry $R\mathcal T$, where $R$ is an orthogonal in-plane spatial operation, these parities and a change of integration variable give
\begin{equation}
 \begin{aligned}
 \alpha_{\mathrm{intra}}&=R\alpha_{\mathrm{intra}}R^T,\\
 \alpha_{\mathrm{inter,s}}&=R\alpha_{\mathrm{inter,s}}R^T,\\
 \alpha_{\mathrm{inter,a}}&=-R\alpha_{\mathrm{inter,a}}R^T.
 \end{aligned}
 \label{eq:appendix-combined-antiunitary}
\end{equation}
The energy-dependent weights and scalar broadening are invariant under the symmetry. In two dimensions, write $\alpha_{\mathrm{inter,a}}=\alpha_{\mathrm a}^{xy}J$, with $J_{xy}=1$, $J_{yx}=-1$, and $J_{xx}=J_{yy}=0$. The identity $RJR^T=(\det R)J$ shows that $C_{2z}\mathcal T$ and $C_{nz}\mathcal T$ forbid this channel, whereas $\mathcal M_x\mathcal T$ permits it. For the symmetric channels, $C_{nz}\mathcal T$ with $n>2$ enforces isotropy, $C_{2z}\mathcal T$ imposes no additional constraint, and $\mathcal M_x\mathcal T$ forbids the transverse component in mirror-aligned axes. Spatial inversion acts as $R=-I$ on the in-plane polar vectors, so $\mathcal{PT}$ has the same channel constraints as $\mathcal T$.

These microscopic parities, together with Eqs.~\eqref{eq:appendix-rotation-invariance} and \eqref{eq:appendix-mirror-invariance}, give all entries in Table~\ref{tab:general-symmetry-main}, including the constraints from combined spatial and time-reversal symmetries.

\subsection{Combined lattice and spin rotations}
\label{app:combined-lattice-spin}

Let $R_6$ rotate momentum counterclockwise by $60^\circ$. It permutes the bond coefficients as $(C_1,C_2,C_3)\mapsto(C_2,C_3,C_1)$, where $C_\ell=C_{\uparrow\downarrow,\ell}$.
For equal-amplitude chiral bond order $(C_{\uparrow\downarrow,1},C_{\uparrow\downarrow,2},C_{\uparrow\downarrow,3})=C(1,\zeta,\zeta^2)$, where $\zeta=e^{2\pi i/3}$, the rotated field is $\zeta\Delta_{\uparrow\downarrow}(\bm k)$. If $\xi=0$ and the diagonal terms retain $C_6$ symmetry, the full Hamiltonian obeys
\begin{equation}
 \begin{aligned}
 h_{\mathrm{MF}}(R_6\bm k)&=S_z h_{\mathrm{MF}}(\bm k)S_z^\dagger,\\
 S_z&=e^{-i(2\pi/3)\sigma_z/2}.
 \end{aligned}
 \label{eq:appendix-compensated-sixfold}
\end{equation}
The opposite chirality uses $S_z^\dagger$. Thus a compensating spin rotation restores the Hamiltonian after a lattice rotation. Charge current and the temperature gradient carry no spin index, so their response satisfies $\alpha=R_6\alpha R_6^T$. Each symmetric channel then has $\alpha^{xx}=\alpha^{yy}$ and $\alpha_{\mathrm s}^{xy}=0$, although the pure lattice rotation alone changes the spin-flip field.

Unequal bond magnitudes or an onsite spin-flip field generally invalidate this particular combined symmetry; the symmetry of the complete Hamiltonian must then be checked. Inversion remains because all bond harmonics are even in momentum. Directionally equivalent, phase-aligned bond fields instead preserve the pure lattice $C_6$ symmetry. These distinctions are needed when interpreting the low-$U$ chiral, intermediate coexistence, and high-$U$ limits.

\section{Triangular-lattice model: mean fields and checks}
\label{app:model-checks-materials}

\subsection{HF coefficients and self-consistency}
\label{app:triangular-hf-symmetry}
\label{app:paramagnetic-metal-check}

Applying the Wick decomposition in Appendix~\ref{app:hartree-fock-details} gives the onsite and nearest-neighbor contributions
\begin{equation}
 \begin{aligned}
 \Sigma_U&=U\begin{pmatrix}n_\downarrow&-\xi^*\\-\xi&n_\uparrow\end{pmatrix},\\
 [\Sigma_V(\bm k)]_{\sigma\sigma'}
 &=12Vn\,\delta_{\sigma\sigma'}-2V\Delta_{\sigma\sigma'}^*(\bm k).
 \end{aligned}
 \label{eq:appendix-triangular-hf-coefficients}
\end{equation}
The nearest-neighbor interaction yields a Hartree shift $12Vn$. Inversion removes the sine harmonics from the exchange term, leaving $-2V\Delta_{\sigma\sigma'}^*$. Adding the bare dispersion and using $n_{\uparrow,\downarrow}=(n\pm m)/2$ yields Eq.~\eqref{eq:triangular-HMF-main}.

Self-consistency follows by evaluating the onsite and cosine-weighted bond averages with $\langle c_{\bm k\sigma}^\dagger c_{\bm k\sigma'}\rangle=\sum_nW_{\sigma n}^*W_{\sigma'n}f_{\bm k n}$. The chemical potential fixes $N^{-1}\sum_{\bm k,n}f_{\bm k n}=n$.

In a fixed spin basis, the classification uses the $z$-component magnetization $m=n_\uparrow-n_\downarrow$ and spin-flip bond fields $C_\ell=C_{\uparrow\downarrow,\ell}$.
Chiral bond order requires at least one pair of nonzero bond fields for which $\arg(C_{\ell'}C_\ell^*)$ differs from both $0$ and $\pi$ modulo $2\pi$.
FM corresponds to $m\ne0$ without chiral bond order, Chiral to $m=0$ with chiral bond order, and Mixed to $m\ne0$ coexisting with chiral bond order.
A solution is classified as a paramagnetic metal (PM) when $m=\xi=C_{\uparrow\downarrow,\ell}=0$ and $C_{\uparrow\uparrow,\ell}=C_{\downarrow\downarrow,\ell}=C$ for every bond direction, with spin-degenerate bands crossing the Fermi level.

\subsection{Checks in limiting cases}

At $V=0$, the spin-dependent part of the HF matrix is $-U(\operatorname{Re}\xi,\operatorname{Im}\xi,m/2)\cdot\bm\sigma$. Its eigenvalue separation is $UM_{\mathrm{on}}$, where $M_{\mathrm{on}}=\sqrt{m^2+4|\xi|^2}$ is the onsite magnetization magnitude. For $n=1.5$, $U/t=8$, and $k_{\mathrm B}T/t=0.01$, the numerical solution reproduces the saturated value $M_{\mathrm{on}}=0.5$ and the corresponding splitting $4t$. This uniform FM solution is consistent with the ferromagnetic state found at the corresponding filling and interaction strength in Ref.~\cite{PasrijaKumar2016}. Since the spin-dependent part is momentum independent, the nondegenerate band eigenstates can be chosen independent of $\bm k$, giving vanishing Berry curvature and quantum metric.

In the homogeneous Hartree limit with $U=0$ and no bond Fock terms, each site of the triangular lattice has six nearest neighbors. With the directed-bond convention used here, each nearest-neighbor bond is counted twice, so all single-particle energies acquire the same shift $12Vn$. At fixed filling and temperature, $\mu$ shifts by the same amount, leaving the energies relative to $\mu$ and the Bloch states unchanged. For $n=1.5$, our numerical scan over $0\leq V/t\leq10$ reproduces $\mu(V)-\mu(0)=12Vn=18V$. After matching the bond-counting convention, this relation agrees with the homogeneous Hartree result in Ref.~\cite{Alekseev2025}.

\subsection{Material-model comparison}

Comparison with a material requires a consistent low-energy band model, hopping amplitudes, and screened onsite and intersite interactions. Constrained random-phase approximation and screened-interaction models provide starting points for such parameter estimates \cite{NakamuraOrganic2009,NakamuraOrganic2012,HansmannSurface2013,TanWigner2023}. The retained band subspace, dimensionality, filling, and orbital or spin--valley degrees of freedom must be matched before comparing a numerical interaction scan with a particular system. Competing orders must also be considered \cite{GneistTriangular2022}. In tunable moir\'e systems \cite{TangMoire2020}, dielectric screening generally changes both onsite and intersite interactions \cite{TanWigner2023}.

\interlinepenalty=10000
\bibliography{references}

\begin{thebibliography}{51}%
\makeatletter
\providecommand \@ifxundefined [1]{%
 \@ifx{#1\undefined}
}%
\providecommand \@ifnum [1]{%
 \ifnum #1\expandafter \@firstoftwo
 \else \expandafter \@secondoftwo
 \fi
}%
\providecommand \@ifx [1]{%
 \ifx #1\expandafter \@firstoftwo
 \else \expandafter \@secondoftwo
 \fi
}%
\providecommand \natexlab [1]{#1}%
\providecommand \enquote  [1]{``#1''}%
\providecommand \bibnamefont  [1]{#1}%
\providecommand \bibfnamefont [1]{#1}%
\providecommand \citenamefont [1]{#1}%
\providecommand \href@noop [0]{\@secondoftwo}%
\providecommand \href [0]{\begingroup \@sanitize@url \@href}%
\providecommand \@href[1]{\@@startlink{#1}\@@href}%
\providecommand \@@href[1]{\endgroup#1\@@endlink}%
\providecommand \@sanitize@url [0]{\catcode `\\12\catcode `\$12\catcode
  `\&12\catcode `\#12\catcode `\^12\catcode `\_12\catcode `\%12\relax}%
\providecommand \@@startlink[1]{}%
\providecommand \@@endlink[0]{}%
\providecommand \url  [0]{\begingroup\@sanitize@url \@url }%
\providecommand \@url [1]{\endgroup\@href {#1}{\urlprefix }}%
\providecommand \urlprefix  [0]{URL }%
\providecommand \Eprint [0]{\href }%
\providecommand \doibase [0]{https://doi.org/}%
\providecommand \selectlanguage [0]{\@gobble}%
\providecommand \bibinfo  [0]{\@secondoftwo}%
\providecommand \bibfield  [0]{\@secondoftwo}%
\providecommand \translation [1]{[#1]}%
\providecommand \BibitemOpen [0]{}%
\providecommand \bibitemStop [0]{}%
\providecommand \bibitemNoStop [0]{.\EOS\space}%
\providecommand \EOS [0]{\spacefactor3000\relax}%
\providecommand \BibitemShut  [1]{\csname bibitem#1\endcsname}%
\let\auto@bib@innerbib\@empty
\bibitem [{\citenamefont {Nagaosa}\ \emph {et~al.}(2010)\citenamefont
  {Nagaosa}, \citenamefont {Sinova}, \citenamefont {Onoda}, \citenamefont
  {MacDonald},\ and\ \citenamefont {Ong}}]{Nagaosa2010}%
  \BibitemOpen
  \bibfield  {author} {\bibinfo {author} {\bibfnamefont {N.}~\bibnamefont
  {Nagaosa}}, \bibinfo {author} {\bibfnamefont {J.}~\bibnamefont {Sinova}},
  \bibinfo {author} {\bibfnamefont {S.}~\bibnamefont {Onoda}}, \bibinfo
  {author} {\bibfnamefont {A.~H.}\ \bibnamefont {MacDonald}},\ and\ \bibinfo
  {author} {\bibfnamefont {N.~P.}\ \bibnamefont {Ong}},\ }\bibfield  {title}
  {\bibinfo {title} {Anomalous {Hall} effect},\ }\href
  {https://doi.org/10.1103/RevModPhys.82.1539} {\bibfield  {journal} {\bibinfo
  {journal} {Rev. Mod. Phys.}\ }\textbf {\bibinfo {volume} {82}},\ \bibinfo
  {pages} {1539} (\bibinfo {year} {2010})}\BibitemShut {NoStop}%
\bibitem [{\citenamefont {Xiao}\ \emph {et~al.}(2006)\citenamefont {Xiao},
  \citenamefont {Yao}, \citenamefont {Fang},\ and\ \citenamefont
  {Niu}}]{Xiao2006}%
  \BibitemOpen
  \bibfield  {author} {\bibinfo {author} {\bibfnamefont {D.}~\bibnamefont
  {Xiao}}, \bibinfo {author} {\bibfnamefont {Y.}~\bibnamefont {Yao}}, \bibinfo
  {author} {\bibfnamefont {Z.}~\bibnamefont {Fang}},\ and\ \bibinfo {author}
  {\bibfnamefont {Q.}~\bibnamefont {Niu}},\ }\bibfield  {title} {\bibinfo
  {title} {{Berry}-phase effect in anomalous thermoelectric transport},\ }\href
  {https://doi.org/10.1103/PhysRevLett.97.026603} {\bibfield  {journal}
  {\bibinfo  {journal} {Phys. Rev. Lett.}\ }\textbf {\bibinfo {volume} {97}},\
  \bibinfo {pages} {026603} (\bibinfo {year} {2006})}\BibitemShut {NoStop}%
\bibitem [{\citenamefont {Xiao}\ \emph {et~al.}(2010)\citenamefont {Xiao},
  \citenamefont {Chang},\ and\ \citenamefont {Niu}}]{Xiao2010}%
  \BibitemOpen
  \bibfield  {author} {\bibinfo {author} {\bibfnamefont {D.}~\bibnamefont
  {Xiao}}, \bibinfo {author} {\bibfnamefont {M.-C.}\ \bibnamefont {Chang}},\
  and\ \bibinfo {author} {\bibfnamefont {Q.}~\bibnamefont {Niu}},\ }\bibfield
  {title} {\bibinfo {title} {{Berry} phase effects on electronic properties},\
  }\href {https://doi.org/10.1103/RevModPhys.82.1959} {\bibfield  {journal}
  {\bibinfo  {journal} {Rev. Mod. Phys.}\ }\textbf {\bibinfo {volume} {82}},\
  \bibinfo {pages} {1959} (\bibinfo {year} {2010})}\BibitemShut {NoStop}%
\bibitem [{\citenamefont {Noky}\ \emph {et~al.}(2018)\citenamefont {Noky},
  \citenamefont {Gooth}, \citenamefont {Felser},\ and\ \citenamefont
  {Sun}}]{Noky2018ANE}%
  \BibitemOpen
  \bibfield  {author} {\bibinfo {author} {\bibfnamefont {J.}~\bibnamefont
  {Noky}}, \bibinfo {author} {\bibfnamefont {J.}~\bibnamefont {Gooth}},
  \bibinfo {author} {\bibfnamefont {C.}~\bibnamefont {Felser}},\ and\ \bibinfo
  {author} {\bibfnamefont {Y.}~\bibnamefont {Sun}},\ }\bibfield  {title}
  {\bibinfo {title} {Characterization of topological band structures away from
  the {Fermi} level by the anomalous {Nernst} effect},\ }\href
  {https://doi.org/10.1103/PhysRevB.98.241106} {\bibfield  {journal} {\bibinfo
  {journal} {Phys. Rev. B}\ }\textbf {\bibinfo {volume} {98}},\ \bibinfo
  {pages} {241106(R)} (\bibinfo {year} {2018})}\BibitemShut {NoStop}%
\bibitem [{\citenamefont {Minami}\ \emph {et~al.}(2020)\citenamefont {Minami},
  \citenamefont {Ishii}, \citenamefont {Hirayama}, \citenamefont {Nomoto},
  \citenamefont {Koretsune},\ and\ \citenamefont
  {Arita}}]{Minami2020NodalLines}%
  \BibitemOpen
  \bibfield  {author} {\bibinfo {author} {\bibfnamefont {S.}~\bibnamefont
  {Minami}}, \bibinfo {author} {\bibfnamefont {F.}~\bibnamefont {Ishii}},
  \bibinfo {author} {\bibfnamefont {M.}~\bibnamefont {Hirayama}}, \bibinfo
  {author} {\bibfnamefont {T.}~\bibnamefont {Nomoto}}, \bibinfo {author}
  {\bibfnamefont {T.}~\bibnamefont {Koretsune}},\ and\ \bibinfo {author}
  {\bibfnamefont {R.}~\bibnamefont {Arita}},\ }\bibfield  {title} {\bibinfo
  {title} {Enhancement of the transverse thermoelectric conductivity
  originating from stationary points in nodal lines},\ }\href
  {https://doi.org/10.1103/PhysRevB.102.205128} {\bibfield  {journal} {\bibinfo
   {journal} {Phys. Rev. B}\ }\textbf {\bibinfo {volume} {102}},\ \bibinfo
  {pages} {205128} (\bibinfo {year} {2020})}\BibitemShut {NoStop}%
\bibitem [{\citenamefont {Wen}\ \emph {et~al.}(2025)\citenamefont {Wen},
  \citenamefont {Xie}, \citenamefont {Auerbach},\ and\ \citenamefont
  {Uchoa}}]{WenXieAuerbachUchoa2025}%
  \BibitemOpen
  \bibfield  {author} {\bibinfo {author} {\bibfnamefont {K.}~\bibnamefont
  {Wen}}, \bibinfo {author} {\bibfnamefont {H.-Y.}\ \bibnamefont {Xie}},
  \bibinfo {author} {\bibfnamefont {A.}~\bibnamefont {Auerbach}},\ and\
  \bibinfo {author} {\bibfnamefont {B.}~\bibnamefont {Uchoa}},\ }\bibfield
  {title} {\bibinfo {title} {Thermal and thermoelectric transport in flat bands
  with nontrivial quantum geometry},\ }\href
  {https://doi.org/10.1103/PhysRevB.111.205140} {\bibfield  {journal} {\bibinfo
   {journal} {Phys. Rev. B}\ }\textbf {\bibinfo {volume} {111}},\ \bibinfo
  {pages} {205140} (\bibinfo {year} {2025})}\BibitemShut {NoStop}%
\bibitem [{\citenamefont {Wen}\ \emph {et~al.}(2026)\citenamefont {Wen},
  \citenamefont {Xie}, \citenamefont {Auerbach},\ and\ \citenamefont
  {Uchoa}}]{WenErratum2026}%
  \BibitemOpen
  \bibfield  {author} {\bibinfo {author} {\bibfnamefont {K.}~\bibnamefont
  {Wen}}, \bibinfo {author} {\bibfnamefont {H.-Y.}\ \bibnamefont {Xie}},
  \bibinfo {author} {\bibfnamefont {A.}~\bibnamefont {Auerbach}},\ and\
  \bibinfo {author} {\bibfnamefont {B.}~\bibnamefont {Uchoa}},\ }\bibfield
  {title} {\bibinfo {title} {Erratum: Thermal and thermoelectric transport in
  flat bands with nontrivial quantum geometry [{Phys. Rev. B} 111, 205140
  (2025)]},\ }\href {https://doi.org/10.1103/2jlf-58dt} {\bibfield  {journal}
  {\bibinfo  {journal} {Phys. Rev. B}\ }\textbf {\bibinfo {volume} {114}},\
  \bibinfo {pages} {049905} (\bibinfo {year} {2026})}\BibitemShut {NoStop}%
\bibitem [{\citenamefont {Lhachemi}\ and\ \citenamefont
  {Cano}(2026)}]{LhachemiCano2026}%
  \BibitemOpen
  \bibfield  {author} {\bibinfo {author} {\bibfnamefont {M.~N.~Y.}\
  \bibnamefont {Lhachemi}}\ and\ \bibinfo {author} {\bibfnamefont
  {J.}~\bibnamefont {Cano}},\ }\href@noop {} {\bibinfo {title} {A unifying
  framework for sum rules and bounds on optical, thermoelectric and thermal
  transport from quantum geometry}} (\bibinfo {year} {2026}),\ \Eprint
  {https://arxiv.org/abs/2603.10121} {arXiv:2603.10121 [cond-mat.mes-hall]}
  \BibitemShut {NoStop}%
\bibitem [{\citenamefont {Sakai}\ \emph {et~al.}(2018)\citenamefont {Sakai},
  \citenamefont {Mizuta}, \citenamefont {Nugroho}, \citenamefont {Sihombing},
  \citenamefont {Koretsune}, \citenamefont {Suzuki}, \citenamefont {Takemori},
  \citenamefont {Ishii}, \citenamefont {Nishio-Hamane}, \citenamefont {Arita},
  \citenamefont {Goswami},\ and\ \citenamefont {Nakatsuji}}]{Sakai2018Co2MnGa}%
  \BibitemOpen
  \bibfield  {author} {\bibinfo {author} {\bibfnamefont {A.}~\bibnamefont
  {Sakai}}, \bibinfo {author} {\bibfnamefont {Y.~P.}\ \bibnamefont {Mizuta}},
  \bibinfo {author} {\bibfnamefont {A.~A.}\ \bibnamefont {Nugroho}}, \bibinfo
  {author} {\bibfnamefont {R.}~\bibnamefont {Sihombing}}, \bibinfo {author}
  {\bibfnamefont {T.}~\bibnamefont {Koretsune}}, \bibinfo {author}
  {\bibfnamefont {M.-T.}\ \bibnamefont {Suzuki}}, \bibinfo {author}
  {\bibfnamefont {N.}~\bibnamefont {Takemori}}, \bibinfo {author}
  {\bibfnamefont {R.}~\bibnamefont {Ishii}}, \bibinfo {author} {\bibfnamefont
  {D.}~\bibnamefont {Nishio-Hamane}}, \bibinfo {author} {\bibfnamefont
  {R.}~\bibnamefont {Arita}}, \bibinfo {author} {\bibfnamefont
  {P.}~\bibnamefont {Goswami}},\ and\ \bibinfo {author} {\bibfnamefont
  {S.}~\bibnamefont {Nakatsuji}},\ }\bibfield  {title} {\bibinfo {title} {Giant
  anomalous {Nernst} effect and quantum-critical scaling in a ferromagnetic
  semimetal},\ }\href {https://doi.org/10.1038/s41567-018-0225-6} {\bibfield
  {journal} {\bibinfo  {journal} {Nat. Phys.}\ }\textbf {\bibinfo {volume}
  {14}},\ \bibinfo {pages} {1119} (\bibinfo {year} {2018})}\BibitemShut
  {NoStop}%
\bibitem [{\citenamefont {Ikhlas}\ \emph {et~al.}(2017)\citenamefont {Ikhlas},
  \citenamefont {Tomita}, \citenamefont {Koretsune}, \citenamefont {Suzuki},
  \citenamefont {Nishio-Hamane}, \citenamefont {Arita}, \citenamefont {Otani},\
  and\ \citenamefont {Nakatsuji}}]{Ikhlas2017Mn3Sn}%
  \BibitemOpen
  \bibfield  {author} {\bibinfo {author} {\bibfnamefont {M.}~\bibnamefont
  {Ikhlas}}, \bibinfo {author} {\bibfnamefont {T.}~\bibnamefont {Tomita}},
  \bibinfo {author} {\bibfnamefont {T.}~\bibnamefont {Koretsune}}, \bibinfo
  {author} {\bibfnamefont {M.-T.}\ \bibnamefont {Suzuki}}, \bibinfo {author}
  {\bibfnamefont {D.}~\bibnamefont {Nishio-Hamane}}, \bibinfo {author}
  {\bibfnamefont {R.}~\bibnamefont {Arita}}, \bibinfo {author} {\bibfnamefont
  {Y.}~\bibnamefont {Otani}},\ and\ \bibinfo {author} {\bibfnamefont
  {S.}~\bibnamefont {Nakatsuji}},\ }\bibfield  {title} {\bibinfo {title} {Large
  anomalous {Nernst} effect at room temperature in a chiral antiferromagnet},\
  }\href {https://doi.org/10.1038/nphys4181} {\bibfield  {journal} {\bibinfo
  {journal} {Nat. Phys.}\ }\textbf {\bibinfo {volume} {13}},\ \bibinfo {pages}
  {1085} (\bibinfo {year} {2017})}\BibitemShut {NoStop}%
\bibitem [{\citenamefont {Li}\ \emph {et~al.}(2017)\citenamefont {Li},
  \citenamefont {Xu}, \citenamefont {Ding}, \citenamefont {Wang}, \citenamefont
  {Shen}, \citenamefont {Lu}, \citenamefont {Zhu},\ and\ \citenamefont
  {Behnia}}]{Li2017Mn3Sn}%
  \BibitemOpen
  \bibfield  {author} {\bibinfo {author} {\bibfnamefont {X.}~\bibnamefont
  {Li}}, \bibinfo {author} {\bibfnamefont {L.}~\bibnamefont {Xu}}, \bibinfo
  {author} {\bibfnamefont {L.}~\bibnamefont {Ding}}, \bibinfo {author}
  {\bibfnamefont {J.}~\bibnamefont {Wang}}, \bibinfo {author} {\bibfnamefont
  {M.}~\bibnamefont {Shen}}, \bibinfo {author} {\bibfnamefont {X.}~\bibnamefont
  {Lu}}, \bibinfo {author} {\bibfnamefont {Z.}~\bibnamefont {Zhu}},\ and\
  \bibinfo {author} {\bibfnamefont {K.}~\bibnamefont {Behnia}},\ }\bibfield
  {title} {\bibinfo {title} {Anomalous {Nernst} and {Righi-Leduc} effects in
  {Mn$_3$Sn}: {Berry} curvature and entropy flow},\ }\href
  {https://doi.org/10.1103/PhysRevLett.119.056601} {\bibfield  {journal}
  {\bibinfo  {journal} {Phys. Rev. Lett.}\ }\textbf {\bibinfo {volume} {119}},\
  \bibinfo {pages} {056601} (\bibinfo {year} {2017})}\BibitemShut {NoStop}%
\bibitem [{\citenamefont {Li}\ \emph {et~al.}(2023)\citenamefont {Li},
  \citenamefont {Zhou}, \citenamefont {Li}, \citenamefont {Qiao}, \citenamefont
  {Jiang}, \citenamefont {Chen}, \citenamefont {Li}, \citenamefont {Tao},\ and\
  \citenamefont {Xu}}]{Li2023Fe3Sn2}%
  \BibitemOpen
  \bibfield  {author} {\bibinfo {author} {\bibfnamefont {Y.}~\bibnamefont
  {Li}}, \bibinfo {author} {\bibfnamefont {J.}~\bibnamefont {Zhou}}, \bibinfo
  {author} {\bibfnamefont {M.}~\bibnamefont {Li}}, \bibinfo {author}
  {\bibfnamefont {L.}~\bibnamefont {Qiao}}, \bibinfo {author} {\bibfnamefont
  {C.}~\bibnamefont {Jiang}}, \bibinfo {author} {\bibfnamefont
  {Q.}~\bibnamefont {Chen}}, \bibinfo {author} {\bibfnamefont {Y.}~\bibnamefont
  {Li}}, \bibinfo {author} {\bibfnamefont {Q.}~\bibnamefont {Tao}},\ and\
  \bibinfo {author} {\bibfnamefont {Z.-A.}\ \bibnamefont {Xu}},\ }\bibfield
  {title} {\bibinfo {title} {Enhanced anomalous {Nernst} effect by tuning the
  chemical potential in the topological kagome ferromagnet {Fe$_3$Sn$_2$}},\
  }\href {https://doi.org/10.1103/PhysRevApplied.19.014026} {\bibfield
  {journal} {\bibinfo  {journal} {Phys. Rev. Applied}\ }\textbf {\bibinfo
  {volume} {19}},\ \bibinfo {pages} {014026} (\bibinfo {year}
  {2023})}\BibitemShut {NoStop}%
\bibitem [{\citenamefont {Guin}\ \emph {et~al.}(2019)\citenamefont {Guin},
  \citenamefont {Vir}, \citenamefont {Zhang}, \citenamefont {Kumar},
  \citenamefont {Watzman}, \citenamefont {Fu}, \citenamefont {Liu},
  \citenamefont {Manna}, \citenamefont {Schnelle}, \citenamefont {Gooth},
  \citenamefont {Shekhar}, \citenamefont {Sun},\ and\ \citenamefont
  {Felser}}]{Guin2019Co3Sn2S2}%
  \BibitemOpen
  \bibfield  {author} {\bibinfo {author} {\bibfnamefont {S.~N.}\ \bibnamefont
  {Guin}}, \bibinfo {author} {\bibfnamefont {P.}~\bibnamefont {Vir}}, \bibinfo
  {author} {\bibfnamefont {Y.}~\bibnamefont {Zhang}}, \bibinfo {author}
  {\bibfnamefont {N.}~\bibnamefont {Kumar}}, \bibinfo {author} {\bibfnamefont
  {S.~J.}\ \bibnamefont {Watzman}}, \bibinfo {author} {\bibfnamefont
  {C.}~\bibnamefont {Fu}}, \bibinfo {author} {\bibfnamefont {E.}~\bibnamefont
  {Liu}}, \bibinfo {author} {\bibfnamefont {K.}~\bibnamefont {Manna}}, \bibinfo
  {author} {\bibfnamefont {W.}~\bibnamefont {Schnelle}}, \bibinfo {author}
  {\bibfnamefont {J.}~\bibnamefont {Gooth}}, \bibinfo {author} {\bibfnamefont
  {C.}~\bibnamefont {Shekhar}}, \bibinfo {author} {\bibfnamefont
  {Y.}~\bibnamefont {Sun}},\ and\ \bibinfo {author} {\bibfnamefont
  {C.}~\bibnamefont {Felser}},\ }\bibfield  {title} {\bibinfo {title}
  {Zero-field {Nernst} effect in a ferromagnetic kagome-lattice
  {Weyl}-semimetal {Co$_3$Sn$_2$S$_2$}},\ }\href
  {https://doi.org/10.1002/adma.201806622} {\bibfield  {journal} {\bibinfo
  {journal} {Adv. Mater.}\ }\textbf {\bibinfo {volume} {31}},\ \bibinfo {pages}
  {1806622} (\bibinfo {year} {2019})}\BibitemShut {NoStop}%
\bibitem [{\citenamefont {Yang}\ \emph {et~al.}(2020)\citenamefont {Yang},
  \citenamefont {You}, \citenamefont {Wang}, \citenamefont {Huang},
  \citenamefont {Xi}, \citenamefont {Xu}, \citenamefont {Cao}, \citenamefont
  {Tian}, \citenamefont {Xu}, \citenamefont {Dai},\ and\ \citenamefont
  {Li}}]{Yang2020}%
  \BibitemOpen
  \bibfield  {author} {\bibinfo {author} {\bibfnamefont {H.}~\bibnamefont
  {Yang}}, \bibinfo {author} {\bibfnamefont {W.}~\bibnamefont {You}}, \bibinfo
  {author} {\bibfnamefont {J.}~\bibnamefont {Wang}}, \bibinfo {author}
  {\bibfnamefont {J.}~\bibnamefont {Huang}}, \bibinfo {author} {\bibfnamefont
  {C.}~\bibnamefont {Xi}}, \bibinfo {author} {\bibfnamefont {X.}~\bibnamefont
  {Xu}}, \bibinfo {author} {\bibfnamefont {C.}~\bibnamefont {Cao}}, \bibinfo
  {author} {\bibfnamefont {M.}~\bibnamefont {Tian}}, \bibinfo {author}
  {\bibfnamefont {Z.-A.}\ \bibnamefont {Xu}}, \bibinfo {author} {\bibfnamefont
  {J.}~\bibnamefont {Dai}},\ and\ \bibinfo {author} {\bibfnamefont
  {Y.}~\bibnamefont {Li}},\ }\bibfield  {title} {\bibinfo {title} {Giant
  anomalous {Nernst} effect in the magnetic {Weyl} semimetal
  {Co$_3$Sn$_2$S$_2$}},\ }\href
  {https://doi.org/10.1103/PhysRevMaterials.4.024202} {\bibfield  {journal}
  {\bibinfo  {journal} {Phys. Rev. Materials}\ }\textbf {\bibinfo {volume}
  {4}},\ \bibinfo {pages} {024202} (\bibinfo {year} {2020})}\BibitemShut
  {NoStop}%
\bibitem [{\citenamefont {Xu}\ \emph {et~al.}(2020)\citenamefont {Xu},
  \citenamefont {Zhao}, \citenamefont {Yi}, \citenamefont {Wang}, \citenamefont
  {Yin}, \citenamefont {Wang}, \citenamefont {Hu}, \citenamefont {Wang},
  \citenamefont {Liu}, \citenamefont {Xu}, \citenamefont {Lu}, \citenamefont
  {Soluyanov}, \citenamefont {Lei}, \citenamefont {Shi}, \citenamefont {Luo},\
  and\ \citenamefont {Chen}}]{Xu2020CorrelatedCo3Sn2S2}%
  \BibitemOpen
  \bibfield  {author} {\bibinfo {author} {\bibfnamefont {Y.}~\bibnamefont
  {Xu}}, \bibinfo {author} {\bibfnamefont {J.}~\bibnamefont {Zhao}}, \bibinfo
  {author} {\bibfnamefont {C.}~\bibnamefont {Yi}}, \bibinfo {author}
  {\bibfnamefont {Q.}~\bibnamefont {Wang}}, \bibinfo {author} {\bibfnamefont
  {Q.}~\bibnamefont {Yin}}, \bibinfo {author} {\bibfnamefont {Y.}~\bibnamefont
  {Wang}}, \bibinfo {author} {\bibfnamefont {X.}~\bibnamefont {Hu}}, \bibinfo
  {author} {\bibfnamefont {L.}~\bibnamefont {Wang}}, \bibinfo {author}
  {\bibfnamefont {E.}~\bibnamefont {Liu}}, \bibinfo {author} {\bibfnamefont
  {G.}~\bibnamefont {Xu}}, \bibinfo {author} {\bibfnamefont {L.}~\bibnamefont
  {Lu}}, \bibinfo {author} {\bibfnamefont {A.~A.}\ \bibnamefont {Soluyanov}},
  \bibinfo {author} {\bibfnamefont {H.}~\bibnamefont {Lei}}, \bibinfo {author}
  {\bibfnamefont {Y.}~\bibnamefont {Shi}}, \bibinfo {author} {\bibfnamefont
  {J.}~\bibnamefont {Luo}},\ and\ \bibinfo {author} {\bibfnamefont {Z.-G.}\
  \bibnamefont {Chen}},\ }\bibfield  {title} {\bibinfo {title} {Electronic
  correlations and flattened band in magnetic {Weyl} semimetal candidate
  $\mathrm{Co}_3\mathrm{Sn}_2\mathrm{S}_2$},\ }\href
  {https://doi.org/10.1038/s41467-020-17234-0} {\bibfield  {journal} {\bibinfo
  {journal} {Nat. Commun.}\ }\textbf {\bibinfo {volume} {11}},\ \bibinfo
  {pages} {3985} (\bibinfo {year} {2020})}\BibitemShut {NoStop}%
\bibitem [{\citenamefont {Asaba}\ \emph {et~al.}(2021)\citenamefont {Asaba},
  \citenamefont {Ivanov}, \citenamefont {Thomas}, \citenamefont {Savrasov},
  \citenamefont {Thompson}, \citenamefont {Bauer},\ and\ \citenamefont
  {Ronning}}]{Asaba2021}%
  \BibitemOpen
  \bibfield  {author} {\bibinfo {author} {\bibfnamefont {T.}~\bibnamefont
  {Asaba}}, \bibinfo {author} {\bibfnamefont {V.}~\bibnamefont {Ivanov}},
  \bibinfo {author} {\bibfnamefont {S.~M.}\ \bibnamefont {Thomas}}, \bibinfo
  {author} {\bibfnamefont {S.~Y.}\ \bibnamefont {Savrasov}}, \bibinfo {author}
  {\bibfnamefont {J.~D.}\ \bibnamefont {Thompson}}, \bibinfo {author}
  {\bibfnamefont {E.~D.}\ \bibnamefont {Bauer}},\ and\ \bibinfo {author}
  {\bibfnamefont {F.}~\bibnamefont {Ronning}},\ }\bibfield  {title} {\bibinfo
  {title} {Colossal anomalous {Nernst} effect in a correlated
  noncentrosymmetric kagome ferromagnet},\ }\href
  {https://doi.org/10.1126/sciadv.abf1467} {\bibfield  {journal} {\bibinfo
  {journal} {Sci. Adv.}\ }\textbf {\bibinfo {volume} {7}},\ \bibinfo {pages}
  {eabf1467} (\bibinfo {year} {2021})}\BibitemShut {NoStop}%
\bibitem [{\citenamefont {Cheng}\ \emph {et~al.}(2023)\citenamefont {Cheng},
  \citenamefont {Huang}, \citenamefont {Zheng}, \citenamefont {Chen},
  \citenamefont {Cochran}, \citenamefont {Hu}, \citenamefont {Yin},
  \citenamefont {Yang}, \citenamefont {Hossain}, \citenamefont {Zhang},
  \citenamefont {Belopolski}, \citenamefont {Liu}, \citenamefont {Cheng},
  \citenamefont {Hashimoto}, \citenamefont {Lu}, \citenamefont {Xu},
  \citenamefont {Zhou}, \citenamefont {Ma}, \citenamefont {Chang},
  \citenamefont {Yao}, \citenamefont {Yin}, \citenamefont {Hasan},\ and\
  \citenamefont {Jia}}]{Cheng2023KondoHall}%
  \BibitemOpen
  \bibfield  {author} {\bibinfo {author} {\bibfnamefont {Z.-J.}\ \bibnamefont
  {Cheng}}, \bibinfo {author} {\bibfnamefont {Y.}~\bibnamefont {Huang}},
  \bibinfo {author} {\bibfnamefont {P.}~\bibnamefont {Zheng}}, \bibinfo
  {author} {\bibfnamefont {L.}~\bibnamefont {Chen}}, \bibinfo {author}
  {\bibfnamefont {T.~A.}\ \bibnamefont {Cochran}}, \bibinfo {author}
  {\bibfnamefont {H.}~\bibnamefont {Hu}}, \bibinfo {author} {\bibfnamefont
  {J.-X.}\ \bibnamefont {Yin}}, \bibinfo {author} {\bibfnamefont {X.~P.}\
  \bibnamefont {Yang}}, \bibinfo {author} {\bibfnamefont {M.~S.}\ \bibnamefont
  {Hossain}}, \bibinfo {author} {\bibfnamefont {Q.}~\bibnamefont {Zhang}},
  \bibinfo {author} {\bibfnamefont {I.}~\bibnamefont {Belopolski}}, \bibinfo
  {author} {\bibfnamefont {R.}~\bibnamefont {Liu}}, \bibinfo {author}
  {\bibfnamefont {G.}~\bibnamefont {Cheng}}, \bibinfo {author} {\bibfnamefont
  {M.}~\bibnamefont {Hashimoto}}, \bibinfo {author} {\bibfnamefont
  {D.}~\bibnamefont {Lu}}, \bibinfo {author} {\bibfnamefont {X.}~\bibnamefont
  {Xu}}, \bibinfo {author} {\bibfnamefont {H.}~\bibnamefont {Zhou}}, \bibinfo
  {author} {\bibfnamefont {W.}~\bibnamefont {Ma}}, \bibinfo {author}
  {\bibfnamefont {G.}~\bibnamefont {Chang}}, \bibinfo {author} {\bibfnamefont
  {N.}~\bibnamefont {Yao}}, \bibinfo {author} {\bibfnamefont {Z.}~\bibnamefont
  {Yin}}, \bibinfo {author} {\bibfnamefont {M.~Z.}\ \bibnamefont {Hasan}},\
  and\ \bibinfo {author} {\bibfnamefont {S.}~\bibnamefont {Jia}},\ }\href@noop
  {} {\bibinfo {title} {Observation of {Kondo} lattice and {Kondo}-enhanced
  anomalous {Hall} effect in an itinerant ferromagnet}} (\bibinfo {year}
  {2023}),\ \Eprint {https://arxiv.org/abs/2302.12113} {arXiv:2302.12113
  [cond-mat.str-el]} \BibitemShut {NoStop}%
\bibitem [{\citenamefont {Guan}\ \emph {et~al.}(2026)\citenamefont {Guan},
  \citenamefont {Guo}, \citenamefont {Zheng}, \citenamefont {Lin},
  \citenamefont {Huang}, \citenamefont {Li}, \citenamefont {Qiang},
  \citenamefont {Li}, \citenamefont {Xie}, \citenamefont {Lu}, \citenamefont
  {Yin},\ and\ \citenamefont {Jia}}]{Guan2026}%
  \BibitemOpen
  \bibfield  {author} {\bibinfo {author} {\bibfnamefont {S.}~\bibnamefont
  {Guan}}, \bibinfo {author} {\bibfnamefont {W.}~\bibnamefont {Guo}}, \bibinfo
  {author} {\bibfnamefont {P.}~\bibnamefont {Zheng}}, \bibinfo {author}
  {\bibfnamefont {X.}~\bibnamefont {Lin}}, \bibinfo {author} {\bibfnamefont
  {Y.}~\bibnamefont {Huang}}, \bibinfo {author} {\bibfnamefont
  {J.}~\bibnamefont {Li}}, \bibinfo {author} {\bibfnamefont {X.-B.}\
  \bibnamefont {Qiang}}, \bibinfo {author} {\bibfnamefont {L.}~\bibnamefont
  {Li}}, \bibinfo {author} {\bibfnamefont {W.}~\bibnamefont {Xie}}, \bibinfo
  {author} {\bibfnamefont {H.-Z.}\ \bibnamefont {Lu}}, \bibinfo {author}
  {\bibfnamefont {Z.}~\bibnamefont {Yin}},\ and\ \bibinfo {author}
  {\bibfnamefont {S.}~\bibnamefont {Jia}},\ }\bibfield  {title} {\bibinfo
  {title} {Enhanced anomalous {Nernst} effect in the ferromagnetic {Kondo}
  lattice {CeCo$_2$As$_2$}},\ }\href {https://doi.org/10.1103/52tp-styb}
  {\bibfield  {journal} {\bibinfo  {journal} {Phys. Rev. Lett.}\ }\textbf
  {\bibinfo {volume} {136}},\ \bibinfo {pages} {036505} (\bibinfo {year}
  {2026})}\BibitemShut {NoStop}%
\bibitem [{\citenamefont {Li}\ \emph {et~al.}(2026)\citenamefont {Li},
  \citenamefont {Guan}, \citenamefont {Chi}, \citenamefont {Zhao},
  \citenamefont {Li}, \citenamefont {Lin}, \citenamefont {Xu},\ and\
  \citenamefont {Jia}}]{Li2026}%
  \BibitemOpen
  \bibfield  {author} {\bibinfo {author} {\bibfnamefont {L.}~\bibnamefont
  {Li}}, \bibinfo {author} {\bibfnamefont {S.}~\bibnamefont {Guan}}, \bibinfo
  {author} {\bibfnamefont {S.}~\bibnamefont {Chi}}, \bibinfo {author}
  {\bibfnamefont {J.}~\bibnamefont {Zhao}}, \bibinfo {author} {\bibfnamefont
  {J.}~\bibnamefont {Li}}, \bibinfo {author} {\bibfnamefont {X.}~\bibnamefont
  {Lin}}, \bibinfo {author} {\bibfnamefont {G.}~\bibnamefont {Xu}},\ and\
  \bibinfo {author} {\bibfnamefont {S.}~\bibnamefont {Jia}},\ }\bibfield
  {title} {\bibinfo {title} {Giant anomalous {{Hall}} and {{Nernst}} effects in
  a heavy fermion ferromagnet},\ }\href
  {https://doi.org/10.1016/j.scib.2026.04.027} {\bibfield  {journal} {\bibinfo
  {journal} {Sci. Bull.}\ }\textbf {\bibinfo {volume} {71}},\ \bibinfo {pages}
  {2463} (\bibinfo {year} {2026})}\BibitemShut {NoStop}%
\bibitem [{\citenamefont {Kubo}\ \emph {et~al.}(1957)\citenamefont {Kubo},
  \citenamefont {Yokota},\ and\ \citenamefont
  {Nakajima}}]{KuboYokotaNakajima1957}%
  \BibitemOpen
  \bibfield  {author} {\bibinfo {author} {\bibfnamefont {R.}~\bibnamefont
  {Kubo}}, \bibinfo {author} {\bibfnamefont {M.}~\bibnamefont {Yokota}},\ and\
  \bibinfo {author} {\bibfnamefont {S.}~\bibnamefont {Nakajima}},\ }\bibfield
  {title} {\bibinfo {title} {Statistical-mechanical theory of irreversible
  processes. {II}. {Response} to thermal disturbance},\ }\href
  {https://doi.org/10.1143/JPSJ.12.1203} {\bibfield  {journal} {\bibinfo
  {journal} {J. Phys. Soc. Jpn.}\ }\textbf {\bibinfo {volume} {12}},\ \bibinfo
  {pages} {1203} (\bibinfo {year} {1957})}\BibitemShut {NoStop}%
\bibitem [{\citenamefont {Luttinger}(1964)}]{Luttinger1964}%
  \BibitemOpen
  \bibfield  {author} {\bibinfo {author} {\bibfnamefont {J.~M.}\ \bibnamefont
  {Luttinger}},\ }\bibfield  {title} {\bibinfo {title} {Theory of thermal
  transport coefficients},\ }\href {https://doi.org/10.1103/PhysRev.135.A1505}
  {\bibfield  {journal} {\bibinfo  {journal} {Phys. Rev.}\ }\textbf {\bibinfo
  {volume} {135}},\ \bibinfo {pages} {A1505} (\bibinfo {year}
  {1964})}\BibitemShut {NoStop}%
\bibitem [{\citenamefont {Onoda}\ \emph {et~al.}(2008)\citenamefont {Onoda},
  \citenamefont {Sugimoto},\ and\ \citenamefont {Nagaosa}}]{Onoda2008}%
  \BibitemOpen
  \bibfield  {author} {\bibinfo {author} {\bibfnamefont {S.}~\bibnamefont
  {Onoda}}, \bibinfo {author} {\bibfnamefont {N.}~\bibnamefont {Sugimoto}},\
  and\ \bibinfo {author} {\bibfnamefont {N.}~\bibnamefont {Nagaosa}},\
  }\bibfield  {title} {\bibinfo {title} {Quantum transport theory of anomalous
  electric, thermoelectric, and thermal {Hall} effects in ferromagnets},\
  }\href {https://doi.org/10.1103/PhysRevB.77.165103} {\bibfield  {journal}
  {\bibinfo  {journal} {Phys. Rev. B}\ }\textbf {\bibinfo {volume} {77}},\
  \bibinfo {pages} {165103} (\bibinfo {year} {2008})}\BibitemShut {NoStop}%
\bibitem [{\citenamefont {Cooper}\ \emph {et~al.}(1997)\citenamefont {Cooper},
  \citenamefont {Halperin},\ and\ \citenamefont {Ruzin}}]{Cooper1997}%
  \BibitemOpen
  \bibfield  {author} {\bibinfo {author} {\bibfnamefont {N.~R.}\ \bibnamefont
  {Cooper}}, \bibinfo {author} {\bibfnamefont {B.~I.}\ \bibnamefont
  {Halperin}},\ and\ \bibinfo {author} {\bibfnamefont {I.~M.}\ \bibnamefont
  {Ruzin}},\ }\bibfield  {title} {\bibinfo {title} {Thermoelectric response of
  an interacting two-dimensional electron gas in a quantizing magnetic field},\
  }\href {https://doi.org/10.1103/PhysRevB.55.2344} {\bibfield  {journal}
  {\bibinfo  {journal} {Phys. Rev. B}\ }\textbf {\bibinfo {volume} {55}},\
  \bibinfo {pages} {2344} (\bibinfo {year} {1997})}\BibitemShut {NoStop}%
\bibitem [{\citenamefont {Qin}\ \emph {et~al.}(2011)\citenamefont {Qin},
  \citenamefont {Niu},\ and\ \citenamefont {Shi}}]{QinNiuShi2011}%
  \BibitemOpen
  \bibfield  {author} {\bibinfo {author} {\bibfnamefont {T.}~\bibnamefont
  {Qin}}, \bibinfo {author} {\bibfnamefont {Q.}~\bibnamefont {Niu}},\ and\
  \bibinfo {author} {\bibfnamefont {J.}~\bibnamefont {Shi}},\ }\bibfield
  {title} {\bibinfo {title} {Energy magnetization and the thermal {Hall}
  effect},\ }\href {https://doi.org/10.1103/PhysRevLett.107.236601} {\bibfield
  {journal} {\bibinfo  {journal} {Phys. Rev. Lett.}\ }\textbf {\bibinfo
  {volume} {107}},\ \bibinfo {pages} {236601} (\bibinfo {year}
  {2011})}\BibitemShut {NoStop}%
\bibitem [{\citenamefont {Kapustin}\ and\ \citenamefont
  {Spodyneiko}(2021)}]{KapustinSpodyneiko2021}%
  \BibitemOpen
  \bibfield  {author} {\bibinfo {author} {\bibfnamefont {A.}~\bibnamefont
  {Kapustin}}\ and\ \bibinfo {author} {\bibfnamefont {L.}~\bibnamefont
  {Spodyneiko}},\ }\bibfield  {title} {\bibinfo {title} {Microscopic formulas
  for thermoelectric transport coefficients in lattice systems},\ }\href
  {https://doi.org/10.1103/PhysRevB.104.035150} {\bibfield  {journal} {\bibinfo
   {journal} {Phys. Rev. B}\ }\textbf {\bibinfo {volume} {104}},\ \bibinfo
  {pages} {035150} (\bibinfo {year} {2021})}\BibitemShut {NoStop}%
\bibitem [{\citenamefont {Dumitriu-I.}\ \emph {et~al.}(2024)\citenamefont
  {Dumitriu-I.}, \citenamefont {Deaconu}, \citenamefont {Kazantsev},\ and\
  \citenamefont {Principi}}]{DumitriuIovanescu2024}%
  \BibitemOpen
  \bibfield  {author} {\bibinfo {author} {\bibfnamefont {A.~D.}\ \bibnamefont
  {Dumitriu-I.}}, \bibinfo {author} {\bibfnamefont {D.-A.}\ \bibnamefont
  {Deaconu}}, \bibinfo {author} {\bibfnamefont {A.~E.}\ \bibnamefont
  {Kazantsev}},\ and\ \bibinfo {author} {\bibfnamefont {A.}~\bibnamefont
  {Principi}},\ }\bibfield  {title} {\bibinfo {title} {First-order effect of
  electron-electron interactions on the anomalous {Hall} conductivity of
  massive {Dirac} fermions},\ }\href
  {https://doi.org/10.1103/PhysRevB.109.165429} {\bibfield  {journal} {\bibinfo
   {journal} {Phys. Rev. B}\ }\textbf {\bibinfo {volume} {109}},\ \bibinfo
  {pages} {165429} (\bibinfo {year} {2024})}\BibitemShut {NoStop}%
\bibitem [{\citenamefont {Dumitriu-I.}\ \emph {et~al.}(2026)\citenamefont
  {Dumitriu-I.}, \citenamefont {Liu}, \citenamefont {Kazantsev},\ and\
  \citenamefont {Principi}}]{DumitriuIovanescu2026}%
  \BibitemOpen
  \bibfield  {author} {\bibinfo {author} {\bibfnamefont {A.~D.}\ \bibnamefont
  {Dumitriu-I.}}, \bibinfo {author} {\bibfnamefont {F.}~\bibnamefont {Liu}},
  \bibinfo {author} {\bibfnamefont {A.~E.}\ \bibnamefont {Kazantsev}},\ and\
  \bibinfo {author} {\bibfnamefont {A.}~\bibnamefont {Principi}},\ }\href@noop
  {} {\bibinfo {title} {Anomalous thermoelectric {Hall} response of interacting
  {2D} {Dirac} fermions}} (\bibinfo {year} {2026}),\ \Eprint
  {https://arxiv.org/abs/2603.25705} {arXiv:2603.25705 [cond-mat.mes-hall]}
  \BibitemShut {NoStop}%
\bibitem [{\citenamefont {Jiang}\ \emph {et~al.}(2015)\citenamefont {Jiang},
  \citenamefont {Zhang}, \citenamefont {Zhou},\ and\ \citenamefont
  {Wang}}]{JiangZhangZhouWang2015}%
  \BibitemOpen
  \bibfield  {author} {\bibinfo {author} {\bibfnamefont {K.}~\bibnamefont
  {Jiang}}, \bibinfo {author} {\bibfnamefont {Y.}~\bibnamefont {Zhang}},
  \bibinfo {author} {\bibfnamefont {S.}~\bibnamefont {Zhou}},\ and\ \bibinfo
  {author} {\bibfnamefont {Z.}~\bibnamefont {Wang}},\ }\bibfield  {title}
  {\bibinfo {title} {Chiral spin density wave order on the frustrated honeycomb
  and bilayer triangle lattice {Hubbard} model at half-filling},\ }\href
  {https://doi.org/10.1103/PhysRevLett.114.216402} {\bibfield  {journal}
  {\bibinfo  {journal} {Phys. Rev. Lett.}\ }\textbf {\bibinfo {volume} {114}},\
  \bibinfo {pages} {216402} (\bibinfo {year} {2015})}\BibitemShut {NoStop}%
\bibitem [{\citenamefont {Pan}\ \emph {et~al.}(2020)\citenamefont {Pan},
  \citenamefont {Wu},\ and\ \citenamefont {Das~Sarma}}]{PanWuDasSarma2020}%
  \BibitemOpen
  \bibfield  {author} {\bibinfo {author} {\bibfnamefont {H.}~\bibnamefont
  {Pan}}, \bibinfo {author} {\bibfnamefont {F.}~\bibnamefont {Wu}},\ and\
  \bibinfo {author} {\bibfnamefont {S.}~\bibnamefont {Das~Sarma}},\ }\bibfield
  {title} {\bibinfo {title} {Quantum phase diagram of a {Moir\'{e}-Hubbard}
  model},\ }\href {https://doi.org/10.1103/PhysRevB.102.201104} {\bibfield
  {journal} {\bibinfo  {journal} {Phys. Rev. B}\ }\textbf {\bibinfo {volume}
  {102}},\ \bibinfo {pages} {201104(R)} (\bibinfo {year} {2020})}\BibitemShut
  {NoStop}%
\bibitem [{\citenamefont {Devakul}\ and\ \citenamefont
  {Fu}(2022)}]{DevakulFu2022}%
  \BibitemOpen
  \bibfield  {author} {\bibinfo {author} {\bibfnamefont {T.}~\bibnamefont
  {Devakul}}\ and\ \bibinfo {author} {\bibfnamefont {L.}~\bibnamefont {Fu}},\
  }\bibfield  {title} {\bibinfo {title} {Quantum anomalous {Hall} effect from
  inverted charge transfer gap},\ }\href
  {https://doi.org/10.1103/PhysRevX.12.021031} {\bibfield  {journal} {\bibinfo
  {journal} {Phys. Rev. X}\ }\textbf {\bibinfo {volume} {12}},\ \bibinfo
  {pages} {021031} (\bibinfo {year} {2022})}\BibitemShut {NoStop}%
\bibitem [{\citenamefont {Zhang}\ \emph {et~al.}(2008)\citenamefont {Zhang},
  \citenamefont {Tewari}, \citenamefont {Yakovenko},\ and\ \citenamefont
  {Das~Sarma}}]{ZhangTewariYakovenkoDasSarma2008}%
  \BibitemOpen
  \bibfield  {author} {\bibinfo {author} {\bibfnamefont {C.}~\bibnamefont
  {Zhang}}, \bibinfo {author} {\bibfnamefont {S.}~\bibnamefont {Tewari}},
  \bibinfo {author} {\bibfnamefont {V.~M.}\ \bibnamefont {Yakovenko}},\ and\
  \bibinfo {author} {\bibfnamefont {S.}~\bibnamefont {Das~Sarma}},\ }\bibfield
  {title} {\bibinfo {title} {Anomalous {Nernst} effect from a chiral
  $d$-density-wave state in underdoped cuprate superconductors},\ }\href
  {https://doi.org/10.1103/PhysRevB.78.174508} {\bibfield  {journal} {\bibinfo
  {journal} {Phys. Rev. B}\ }\textbf {\bibinfo {volume} {78}},\ \bibinfo
  {pages} {174508} (\bibinfo {year} {2008})}\BibitemShut {NoStop}%
\bibitem [{\citenamefont {Wang}\ and\ \citenamefont {Zhu}(2026)}]{WangZhu2026}%
  \BibitemOpen
  \bibfield  {author} {\bibinfo {author} {\bibfnamefont {Y.}~\bibnamefont
  {Wang}}\ and\ \bibinfo {author} {\bibfnamefont {Z.-G.}\ \bibnamefont {Zhu}},\
  }\href@noop {} {\bibinfo {title} {Intrinsic violation of the
  {Wiedemann--Franz} law in interacting systems}} (\bibinfo {year} {2026}),\
  \Eprint {https://arxiv.org/abs/2603.11713} {arXiv:2603.11713
  [cond-mat.mes-hall]} \BibitemShut {NoStop}%
\bibitem [{\citenamefont {Tatara}(2015)}]{Tatara2015}%
  \BibitemOpen
  \bibfield  {author} {\bibinfo {author} {\bibfnamefont {G.}~\bibnamefont
  {Tatara}},\ }\bibfield  {title} {\bibinfo {title} {Thermal vector potential
  theory of transport induced by a temperature gradient},\ }\href
  {https://doi.org/10.1103/PhysRevLett.114.196601} {\bibfield  {journal}
  {\bibinfo  {journal} {Phys. Rev. Lett.}\ }\textbf {\bibinfo {volume} {114}},\
  \bibinfo {pages} {196601} (\bibinfo {year} {2015})}\BibitemShut {NoStop}%
\bibitem [{\citenamefont {Wang}\ \emph {et~al.}(2022)\citenamefont {Wang},
  \citenamefont {Zhu},\ and\ \citenamefont {Su}}]{Wang2022b}%
  \BibitemOpen
  \bibfield  {author} {\bibinfo {author} {\bibfnamefont {Y.}~\bibnamefont
  {Wang}}, \bibinfo {author} {\bibfnamefont {Z.-G.}\ \bibnamefont {Zhu}},\ and\
  \bibinfo {author} {\bibfnamefont {G.}~\bibnamefont {Su}},\ }\bibfield
  {title} {\bibinfo {title} {Quantum theory of nonlinear thermal response},\
  }\href {https://doi.org/10.1103/PhysRevB.106.035148} {\bibfield  {journal}
  {\bibinfo  {journal} {Phys. Rev. B}\ }\textbf {\bibinfo {volume} {106}},\
  \bibinfo {pages} {035148} (\bibinfo {year} {2022})}\BibitemShut {NoStop}%
\bibitem [{\citenamefont {Li}\ and\ \citenamefont {Zhu}(2024)}]{Li2024}%
  \BibitemOpen
  \bibfield  {author} {\bibinfo {author} {\bibfnamefont {J.-C.}\ \bibnamefont
  {Li}}\ and\ \bibinfo {author} {\bibfnamefont {Z.-G.}\ \bibnamefont {Zhu}},\
  }\bibfield  {title} {\bibinfo {title} {Intrinsic second-order magnon thermal
  {Hall} effect},\ }\href {https://doi.org/10.1088/1361-648X/ad5bb0} {\bibfield
   {journal} {\bibinfo  {journal} {J. Phys.: Condens. Matter}\ }\textbf
  {\bibinfo {volume} {36}},\ \bibinfo {pages} {395802} (\bibinfo {year}
  {2024})}\BibitemShut {NoStop}%
\bibitem [{\citenamefont {Mitscherling}(2020)}]{Mitscherling2020}%
  \BibitemOpen
  \bibfield  {author} {\bibinfo {author} {\bibfnamefont {J.}~\bibnamefont
  {Mitscherling}},\ }\bibfield  {title} {\bibinfo {title} {Longitudinal and
  anomalous {Hall} conductivity of a general two-band model},\ }\href
  {https://doi.org/10.1103/PhysRevB.102.165151} {\bibfield  {journal} {\bibinfo
   {journal} {Phys. Rev. B}\ }\textbf {\bibinfo {volume} {102}},\ \bibinfo
  {pages} {165151} (\bibinfo {year} {2020})}\BibitemShut {NoStop}%
\bibitem [{\citenamefont {Sticlet}\ \emph {et~al.}(2012)\citenamefont
  {Sticlet}, \citenamefont {Pi\'echon}, \citenamefont {Fuchs}, \citenamefont
  {Kalugin},\ and\ \citenamefont {Simon}}]{Sticlet2012}%
  \BibitemOpen
  \bibfield  {author} {\bibinfo {author} {\bibfnamefont {D.}~\bibnamefont
  {Sticlet}}, \bibinfo {author} {\bibfnamefont {F.}~\bibnamefont {Pi\'echon}},
  \bibinfo {author} {\bibfnamefont {J.-N.}\ \bibnamefont {Fuchs}}, \bibinfo
  {author} {\bibfnamefont {P.}~\bibnamefont {Kalugin}},\ and\ \bibinfo {author}
  {\bibfnamefont {P.}~\bibnamefont {Simon}},\ }\bibfield  {title} {\bibinfo
  {title} {Geometrical engineering of a two-band {Chern} insulator in two
  dimensions with arbitrary topological index},\ }\href
  {https://doi.org/10.1103/PhysRevB.85.165456} {\bibfield  {journal} {\bibinfo
  {journal} {Phys. Rev. B}\ }\textbf {\bibinfo {volume} {85}},\ \bibinfo
  {pages} {165456} (\bibinfo {year} {2012})}\BibitemShut {NoStop}%
\bibitem [{\citenamefont {Kotetes}\ and\ \citenamefont
  {Varelogiannis}(2010)}]{KotetesVarelogiannis2010}%
  \BibitemOpen
  \bibfield  {author} {\bibinfo {author} {\bibfnamefont {P.}~\bibnamefont
  {Kotetes}}\ and\ \bibinfo {author} {\bibfnamefont {G.}~\bibnamefont
  {Varelogiannis}},\ }\bibfield  {title} {\bibinfo {title} {Chirality induced
  tilted-hill giant {Nernst} signal},\ }\href
  {https://doi.org/10.1103/PhysRevLett.104.106404} {\bibfield  {journal}
  {\bibinfo  {journal} {Phys. Rev. Lett.}\ }\textbf {\bibinfo {volume} {104}},\
  \bibinfo {pages} {106404} (\bibinfo {year} {2010})}\BibitemShut {NoStop}%
\bibitem [{\citenamefont {Baym}\ and\ \citenamefont
  {Kadanoff}(1961)}]{BaymKadanoff1961}%
  \BibitemOpen
  \bibfield  {author} {\bibinfo {author} {\bibfnamefont {G.}~\bibnamefont
  {Baym}}\ and\ \bibinfo {author} {\bibfnamefont {L.~P.}\ \bibnamefont
  {Kadanoff}},\ }\bibfield  {title} {\bibinfo {title} {Conservation laws and
  correlation functions},\ }\href {https://doi.org/10.1103/PhysRev.124.287}
  {\bibfield  {journal} {\bibinfo  {journal} {Phys. Rev.}\ }\textbf {\bibinfo
  {volume} {124}},\ \bibinfo {pages} {287} (\bibinfo {year}
  {1961})}\BibitemShut {NoStop}%
\bibitem [{\citenamefont {Kontani}(2003)}]{Kontani2003}%
  \BibitemOpen
  \bibfield  {author} {\bibinfo {author} {\bibfnamefont {H.}~\bibnamefont
  {Kontani}},\ }\bibfield  {title} {\bibinfo {title} {General formula for the
  thermoelectric transport phenomena based on {Fermi} liquid theory:
  Thermoelectric power, {Nernst} coefficient, and thermal conductivity},\
  }\href {https://doi.org/10.1103/PhysRevB.67.014408} {\bibfield  {journal}
  {\bibinfo  {journal} {Phys. Rev. B}\ }\textbf {\bibinfo {volume} {67}},\
  \bibinfo {pages} {014408} (\bibinfo {year} {2003})}\BibitemShut {NoStop}%
\bibitem [{\citenamefont {Martin}\ and\ \citenamefont
  {Batista}(2008)}]{MartinBatista2008}%
  \BibitemOpen
  \bibfield  {author} {\bibinfo {author} {\bibfnamefont {I.}~\bibnamefont
  {Martin}}\ and\ \bibinfo {author} {\bibfnamefont {C.~D.}\ \bibnamefont
  {Batista}},\ }\bibfield  {title} {\bibinfo {title} {Itinerant electron-driven
  chiral magnetic ordering and spontaneous quantum {Hall} effect in triangular
  lattice models},\ }\href {https://doi.org/10.1103/PhysRevLett.101.156402}
  {\bibfield  {journal} {\bibinfo  {journal} {Phys. Rev. Lett.}\ }\textbf
  {\bibinfo {volume} {101}},\ \bibinfo {pages} {156402} (\bibinfo {year}
  {2008})}\BibitemShut {NoStop}%
\bibitem [{\citenamefont {Pasrija}\ and\ \citenamefont
  {Kumar}(2016)}]{PasrijaKumar2016}%
  \BibitemOpen
  \bibfield  {author} {\bibinfo {author} {\bibfnamefont {K.}~\bibnamefont
  {Pasrija}}\ and\ \bibinfo {author} {\bibfnamefont {S.}~\bibnamefont
  {Kumar}},\ }\bibfield  {title} {\bibinfo {title} {Noncollinear and
  noncoplanar magnetic order in the extended {Hubbard} model on anisotropic
  triangular lattice},\ }\href {https://doi.org/10.1103/PhysRevB.93.195110}
  {\bibfield  {journal} {\bibinfo  {journal} {Phys. Rev. B}\ }\textbf {\bibinfo
  {volume} {93}},\ \bibinfo {pages} {195110} (\bibinfo {year}
  {2016})}\BibitemShut {NoStop}%
\bibitem [{\citenamefont {Weischenberg}\ \emph {et~al.}(2013)\citenamefont
  {Weischenberg}, \citenamefont {Freimuth}, \citenamefont {Bl{\"u}gel},\ and\
  \citenamefont {Mokrousov}}]{Weischenberg2013}%
  \BibitemOpen
  \bibfield  {author} {\bibinfo {author} {\bibfnamefont {J.}~\bibnamefont
  {Weischenberg}}, \bibinfo {author} {\bibfnamefont {F.}~\bibnamefont
  {Freimuth}}, \bibinfo {author} {\bibfnamefont {S.}~\bibnamefont
  {Bl{\"u}gel}},\ and\ \bibinfo {author} {\bibfnamefont {Y.}~\bibnamefont
  {Mokrousov}},\ }\bibfield  {title} {\bibinfo {title} {Scattering-independent
  anomalous {Nernst} effect in ferromagnets},\ }\href
  {https://doi.org/10.1103/PhysRevB.87.060406} {\bibfield  {journal} {\bibinfo
  {journal} {Phys. Rev. B}\ }\textbf {\bibinfo {volume} {87}},\ \bibinfo
  {pages} {060406(R)} (\bibinfo {year} {2013})}\BibitemShut {NoStop}%
\bibitem [{\citenamefont {Papaj}\ and\ \citenamefont {Fu}(2021)}]{PapajFu2021}%
  \BibitemOpen
  \bibfield  {author} {\bibinfo {author} {\bibfnamefont {M.}~\bibnamefont
  {Papaj}}\ and\ \bibinfo {author} {\bibfnamefont {L.}~\bibnamefont {Fu}},\
  }\bibfield  {title} {\bibinfo {title} {Enhanced anomalous {Nernst} effect in
  disordered {Dirac} and {Weyl} materials},\ }\href
  {https://doi.org/10.1103/PhysRevB.103.075424} {\bibfield  {journal} {\bibinfo
   {journal} {Phys. Rev. B}\ }\textbf {\bibinfo {volume} {103}},\ \bibinfo
  {pages} {075424} (\bibinfo {year} {2021})}\BibitemShut {NoStop}%
\bibitem [{\citenamefont {Alekseev}\ \emph {et~al.}(2025)\citenamefont
  {Alekseev}, \citenamefont {Cichy},\ and\ \citenamefont
  {Kapcia}}]{Alekseev2025}%
  \BibitemOpen
  \bibfield  {author} {\bibinfo {author} {\bibfnamefont {A.}~\bibnamefont
  {Alekseev}}, \bibinfo {author} {\bibfnamefont {A.}~\bibnamefont {Cichy}},\
  and\ \bibinfo {author} {\bibfnamefont {K.~J.}\ \bibnamefont {Kapcia}},\
  }\bibfield  {title} {\bibinfo {title} {Particle-hole asymmetry and pinball
  liquid in a triangular-lattice extended {Hubbard} model within the mean-field
  approximation},\ }\href {https://doi.org/10.1103/134x-s54f} {\bibfield
  {journal} {\bibinfo  {journal} {Phys. Rev. B}\ }\textbf {\bibinfo {volume}
  {112}},\ \bibinfo {pages} {115155} (\bibinfo {year} {2025})}\BibitemShut
  {NoStop}%
\bibitem [{\citenamefont {Nakamura}\ \emph {et~al.}(2009)\citenamefont
  {Nakamura}, \citenamefont {Yoshimoto}, \citenamefont {Kosugi}, \citenamefont
  {Arita},\ and\ \citenamefont {Imada}}]{NakamuraOrganic2009}%
  \BibitemOpen
  \bibfield  {author} {\bibinfo {author} {\bibfnamefont {K.}~\bibnamefont
  {Nakamura}}, \bibinfo {author} {\bibfnamefont {Y.}~\bibnamefont {Yoshimoto}},
  \bibinfo {author} {\bibfnamefont {T.}~\bibnamefont {Kosugi}}, \bibinfo
  {author} {\bibfnamefont {R.}~\bibnamefont {Arita}},\ and\ \bibinfo {author}
  {\bibfnamefont {M.}~\bibnamefont {Imada}},\ }\bibfield  {title} {\bibinfo
  {title} {Ab initio derivation of low-energy model for {$\kappa$-ET} type
  organic conductors},\ }\href {https://doi.org/10.1143/JPSJ.78.083710}
  {\bibfield  {journal} {\bibinfo  {journal} {J. Phys. Soc. Jpn.}\ }\textbf
  {\bibinfo {volume} {78}},\ \bibinfo {pages} {083710} (\bibinfo {year}
  {2009})}\BibitemShut {NoStop}%
\bibitem [{\citenamefont {Nakamura}\ \emph {et~al.}(2012)\citenamefont
  {Nakamura}, \citenamefont {Yoshimoto},\ and\ \citenamefont
  {Imada}}]{NakamuraOrganic2012}%
  \BibitemOpen
  \bibfield  {author} {\bibinfo {author} {\bibfnamefont {K.}~\bibnamefont
  {Nakamura}}, \bibinfo {author} {\bibfnamefont {Y.}~\bibnamefont
  {Yoshimoto}},\ and\ \bibinfo {author} {\bibfnamefont {M.}~\bibnamefont
  {Imada}},\ }\bibfield  {title} {\bibinfo {title} {Ab initio two-dimensional
  multiband low-energy models of {EtMe$_3$Sb[Pd(dmit)$_2$]$_2$} and
  {$\kappa$-(BEDT-TTF)$_2$Cu(NCS)$_2$} with comparisons to single-band
  models},\ }\href {https://doi.org/10.1103/PhysRevB.86.205117} {\bibfield
  {journal} {\bibinfo  {journal} {Phys. Rev. B}\ }\textbf {\bibinfo {volume}
  {86}},\ \bibinfo {pages} {205117} (\bibinfo {year} {2012})}\BibitemShut
  {NoStop}%
\bibitem [{\citenamefont {Hansmann}\ \emph {et~al.}(2013)\citenamefont
  {Hansmann}, \citenamefont {Vaugier}, \citenamefont {Jiang},\ and\
  \citenamefont {Biermann}}]{HansmannSurface2013}%
  \BibitemOpen
  \bibfield  {author} {\bibinfo {author} {\bibfnamefont {P.}~\bibnamefont
  {Hansmann}}, \bibinfo {author} {\bibfnamefont {L.}~\bibnamefont {Vaugier}},
  \bibinfo {author} {\bibfnamefont {H.}~\bibnamefont {Jiang}},\ and\ \bibinfo
  {author} {\bibfnamefont {S.}~\bibnamefont {Biermann}},\ }\bibfield  {title}
  {\bibinfo {title} {What about {U} on surfaces? {Extended} {Hubbard} models
  for adatom systems from first principles},\ }\href
  {https://doi.org/10.1088/0953-8984/25/9/094005} {\bibfield  {journal}
  {\bibinfo  {journal} {J. Phys.: Condens. Matter}\ }\textbf {\bibinfo {volume}
  {25}},\ \bibinfo {pages} {094005} (\bibinfo {year} {2013})}\BibitemShut
  {NoStop}%
\bibitem [{\citenamefont {Tan}\ \emph {et~al.}(2023)\citenamefont {Tan},
  \citenamefont {Tsang}, \citenamefont {Dobrosavljevi{\'c}},\ and\
  \citenamefont {Rademaker}}]{TanWigner2023}%
  \BibitemOpen
  \bibfield  {author} {\bibinfo {author} {\bibfnamefont {Y.}~\bibnamefont
  {Tan}}, \bibinfo {author} {\bibfnamefont {P.~K.~H.}\ \bibnamefont {Tsang}},
  \bibinfo {author} {\bibfnamefont {V.}~\bibnamefont {Dobrosavljevi{\'c}}},\
  and\ \bibinfo {author} {\bibfnamefont {L.}~\bibnamefont {Rademaker}},\
  }\bibfield  {title} {\bibinfo {title} {Doping a {Wigner-Mott} insulator:
  Exotic charge orders in transition metal dichalcogenide moir{\'e}
  heterobilayers},\ }\href {https://doi.org/10.1103/PhysRevResearch.5.043190}
  {\bibfield  {journal} {\bibinfo  {journal} {Phys. Rev. Research}\ }\textbf
  {\bibinfo {volume} {5}},\ \bibinfo {pages} {043190} (\bibinfo {year}
  {2023})}\BibitemShut {NoStop}%
\bibitem [{\citenamefont {Gneist}\ \emph {et~al.}(2022)\citenamefont {Gneist},
  \citenamefont {Classen},\ and\ \citenamefont
  {Scherer}}]{GneistTriangular2022}%
  \BibitemOpen
  \bibfield  {author} {\bibinfo {author} {\bibfnamefont {N.}~\bibnamefont
  {Gneist}}, \bibinfo {author} {\bibfnamefont {L.}~\bibnamefont {Classen}},\
  and\ \bibinfo {author} {\bibfnamefont {M.~M.}\ \bibnamefont {Scherer}},\
  }\bibfield  {title} {\bibinfo {title} {Competing instabilities of the
  extended {Hubbard} model on the triangular lattice: Truncated-unity
  functional renormalization group and application to moir{\'e} materials},\
  }\href {https://doi.org/10.1103/PhysRevB.106.125141} {\bibfield  {journal}
  {\bibinfo  {journal} {Phys. Rev. B}\ }\textbf {\bibinfo {volume} {106}},\
  \bibinfo {pages} {125141} (\bibinfo {year} {2022})}\BibitemShut {NoStop}%
\bibitem [{\citenamefont {Tang}\ \emph {et~al.}(2020)\citenamefont {Tang},
  \citenamefont {Li}, \citenamefont {Li}, \citenamefont {Xu}, \citenamefont
  {Liu}, \citenamefont {Barmak}, \citenamefont {Watanabe}, \citenamefont
  {Taniguchi}, \citenamefont {MacDonald}, \citenamefont {Shan},\ and\
  \citenamefont {Mak}}]{TangMoire2020}%
  \BibitemOpen
  \bibfield  {author} {\bibinfo {author} {\bibfnamefont {Y.}~\bibnamefont
  {Tang}}, \bibinfo {author} {\bibfnamefont {L.}~\bibnamefont {Li}}, \bibinfo
  {author} {\bibfnamefont {T.}~\bibnamefont {Li}}, \bibinfo {author}
  {\bibfnamefont {Y.}~\bibnamefont {Xu}}, \bibinfo {author} {\bibfnamefont
  {S.}~\bibnamefont {Liu}}, \bibinfo {author} {\bibfnamefont {K.}~\bibnamefont
  {Barmak}}, \bibinfo {author} {\bibfnamefont {K.}~\bibnamefont {Watanabe}},
  \bibinfo {author} {\bibfnamefont {T.}~\bibnamefont {Taniguchi}}, \bibinfo
  {author} {\bibfnamefont {A.~H.}\ \bibnamefont {MacDonald}}, \bibinfo {author}
  {\bibfnamefont {J.}~\bibnamefont {Shan}},\ and\ \bibinfo {author}
  {\bibfnamefont {K.~F.}\ \bibnamefont {Mak}},\ }\bibfield  {title} {\bibinfo
  {title} {Simulation of {Hubbard} model physics in {WSe$_2$/WS$_2$} moir{\'e}
  superlattices},\ }\href {https://doi.org/10.1038/s41586-020-2085-3}
  {\bibfield  {journal} {\bibinfo  {journal} {Nature}\ }\textbf {\bibinfo
  {volume} {579}},\ \bibinfo {pages} {353} (\bibinfo {year}
  {2020})}\BibitemShut {NoStop}%
\end{thebibliography}%

\end{document}